\documentclass[floats,floatfix,aps,12pt]{revtex4}
\usepackage{times} 
\usepackage{amssymb}
\usepackage{amsbsy}
\usepackage{color}
\ifx\pdfoutput\undefined
\usepackage{graphicx}
\else
\usepackage[pdftex]{graphicx,hyperref}
\fi
\newcommand{\mylab}[1]{\label{#1}}
\renewcommand{\vec}[1]{\mathbf{#1}}
\newcommand{\vecg}[1]{\boldsymbol{#1}}
\newcommand{\tens}[1]{\mathbf{\underline{#1}}}

\begin{document}
\title[Thin film evolution equations from dewetting to epitaxial growth]{Thin film evolution equations from (evaporating) dewetting liquid layers to epitaxial growth}
\author{U. Thiele\footnote{homepage: http://www.uwethiele.de}}
\email{u.thiele@lboro.ac.uk}
\affiliation{Department of Mathematical Sciences, 
Loughborough University, Leicestershire LE11 3TU, UK}

\begin{abstract}
  In the present contribution we review basic mathematical results for
  three physical systems involving self-organising solid or liquid
  films at solid surfaces. The films may undergo a structuring process
  by dewetting, evaporation/condensation or epitaxial growth,
  respectively. We highlight similarities and differences of the three
  systems based on the observation that in certain limits all of them
  may be described using models of similar form, i.e., time evolution
  equations for the film thickness profile. Those equations represent
  gradient dynamics characterized by mobility functions and an
  underlying energy functional.

  Two basic steps of mathematical analysis are used to compare the
  different system. First, we discuss the linear stability of
  homogeneous steady states, i.e., flat films; and second the
  systematics of non-trivial steady states, i.e., drop/hole states for
  dewetting films and quantum dot states in epitaxial growth,
  respectively. Our aim is to illustrate that the underlying solution
  structure might be very complex as in the case of epitaxial growth
  but can be better understood when comparing to the much simpler
  results for the dewetting liquid film. We furthermore show that the
  numerical continuation techniques employed can shed some light on
  this structure in a more convenient way than time-stepping methods.

  Finally we discuss that the usage of the employed general formulation
  does not only relate seemingly not related physical systems
  mathematically, but does as well allow to discuss model extensions
  in a more unified way.\\[3ex]
   \textcolor{red}{\large The paper is published in: 
{\it J. Phys.-Cond. Mat.} {\bf 22}, 084019 (2010) and can be obtained at
    \href{http://dx.doi.org/10.1088/0953-8984/22/8/084019}{http://dx.doi.org/10.1088/0953-8984/22/8/084019}
}
\end{abstract}
\maketitle

\section{Introduction}
Structure formation at interfaces and surfaces occurs widely in our
natural and technological environment. The spectrum of related
phenomena ranges from growing dendrites in solidification or
crystallization to budding membranes in the biological cell.
Technological processes based on or affected by interfacial
structuring processes involve, for instance, the sputtering of solid
surfaces (that may roughen), the usage of instabilities in the
epitaxial growth of nano- or quantum-dots, the structuring of
homogeneous liquid or elastic coating layers, the deposition of
structured nano-particle assemblies employing instabilities, and
heat-exchanger technology based on transfer enhancement by surface
waves on falling liquid films. A selection is discussed in
\cite{GhWa08}.

Part of the mentioned structuring processes can be modeled as an
evolution in time of a surface profile. Models are normally based
either on a stochastic microscopic discrete or a deterministic
mesoscopic or macroscopic continuum approach. For an overview of
methods for the modelling of (solid) nanostructures see \cite{Vved04}.
Here, we focus on the continuum approach.  There exists an important
subset of systems that evolve towards an equilibrium state
corresponding to an energetic minimum, i.e., normally these are
relaxational systems without any external forcing.

A continuum description of relaxational systems can often be brought
into the form of time evolution equation(s) for one or several
conserved or non-conserved order parameter fields $\phi(\vec{x},t)$
(cf.~\cite{Lang92}).  A non-conserved field might still follow in part
a conserved dynamics. The dynamics is governed by the underlying
energy functional $F[\phi]$. The simplest form for a time evolution of
a purely dissipative system without any inertial hamiltonian dynamics
corresponds to the gradient dynamics
\begin{equation}
\partial_t \phi \,=\,
\nabla\cdot\left[M_{\mathrm{c}}\nabla\frac{\delta F}{\delta\phi}\right]
\,-\, M_{\mathrm{nc}}\frac{\delta F}{\delta\phi}
\mylab{film}
\end{equation}
with the $M_{\mathrm{c}}(\phi)\ge0$ and $M_{\mathrm{nc}}(\phi)\ge0$
being the mobility functions for the conserved and non-conserved part
of the dynamics, respectively. Here and in the following $\partial_t$
and $\partial_x$ denote partial derivatives w.r.t.~ time and space,
respectively.

A typical example is the Cahn-Hilliard equation describing the
demixing of a binary mixture, i.e., a purely 'conserved dynamics'
($M_{\mathrm{nc}}=0$) \cite{CaHi58,Cahn65,Lang92}. Another example is
the Allen-Cahn equation describing, for instance, the dynamics of the
Ising model in the continuum limit \cite{Lang92}. Multiplying
Eq.~(\ref{film}) by $\delta F/\delta\phi$ and integrating one obtains
after a partial integration
\begin{equation}
\frac{d F[\phi]}{dt} \,=\,-\int M_{\mathrm{c}} \left(\nabla\frac{\delta F}{\delta\phi}\right)^2dV
\,-\, \int M_{\mathrm{nc}}\left(\frac{\delta F}{\delta\phi}\right)^2dV \le 0
\mylab{lyap}
\end{equation}
confirming $F[\phi]$ to be a Lyapunov functional.

In the context of evolving surfaces or interfaces, such an equation
appears in various contexts. We will discus here (i) film thickness
equations for films of non-volatile and volatile liquids on solid
substrates and (ii) surface profile equations for epitaxial growth.

Eq.~(\ref{film}) might describe the evolution of the surface profile
of an evaporating or condensing thin liquid film on a solid substrate
under the influence of capillarity and wettability. In this case, the
function $\phi(\vec{x},t)$ represents the film thickness profile and
the functional $F[\phi]$ is given by
\begin{equation}
F[\phi]\,=\,\int\left[\frac{\gamma}{2}(\nabla \phi)^2 + f(\phi) - \mu\phi\right]dV
\label{eq:en1}
\end{equation}
where $\gamma$ is the liquid-gas surface tension and $f(\phi)$ is a
local free energy, related to the disjoining pressure $\Pi(\phi)$ by
$\Pi=-\partial_\phi f(\phi)$. The term $\mu\phi$ represents an overall
energy bias towards the liquid or the gaseous state. It is the sole
responsible for evaporation/condensation of flat 'bulk' films, i.e.,
films that are thick as compared to the range of the disjoining
pressure. $\mu$ corresponds to a chemical potential.  For details and
specific choices for the disjoining pressure see, e.g.,
Refs.~\cite{deGe85,Isra92,Mitl93,ODB97,PiPo00,KaTh07}.

For Poiseuille flow in the film without slip at the substrate the
mobility for the conserved part is $M_{\mathrm{c}}=\phi^3/3\eta$ where
$\eta$ is the dynamic viscosity. Several slip regimes might be
accounted for by different choices for $M_{\mathrm{c}}(\phi)$
\cite{Fetz05}.  The mobility function for the non-conserved part is
normally assumed to be a constant (see \cite{LGP02}). Note that there
exists an ongoing discussion regarding the form of the non-conserved
part of the dynamics
(cf.~e.g.~\cite{BBD88,OrBa99,SREP01,KKS01,LGP02,BeMe06}).

Eq.~(\ref{film}) with (\ref{eq:en1}) is extensively studied in the
conserved case, i.e., for non-volatile films ($M_{\mathrm{nc}}=0$).
It can easily be derived from the Navier-Stokes equations and
appropriate boundary conditions at the substrate and the free surface
employing a long-wave or lubrication approximation
\cite{ODB97,KaTh07}. Depending on the particular physical situation
studied many different forms for the local energy function $f(\phi)$
are encountered. Beside 'proper' disjoining pressures that model
effective molecular interactions between film and substrate
(wettability)
\cite{DLP60,RuJa74,deGe85,Isra92,Hock93,ShKh98,ESR00,PiPo00,TVNP01}
the equations may as well incorporate other pressures modelling, e.g.,
the influence of an electric field on a film of dielectric liquid
in a capacitor \cite{Lin01,Lin02,VSKB05,JoTh07,JHT08} or films
on homogeneously heated substrates that form structures due to a
long-wave Marangoni instability
\cite{BBD88,OrRo94,GNP94,Oron00,BPT03,ThKn04}.  The latter is
especially interesting because it represents a system that is kept
permanently out of equilibrium but is nevertheless described by a
gradient dynamics. Note that the situation is slightly different in a
closed two-layer system \cite{MPBT05} that can in two dimensions be
described by Eq.~(\ref{film}) with an appropriate $F[\phi]$ but not in
three dimensions.

The above mentioned Cahn-Hilliard equation describing the conserved
dynamics of demixing of a binary mixture corresponds to
Eq.~(\ref{film}) and Eq.~(\ref{eq:en1}) with a constant
$M_{\mathrm{c}}$, $M_{\mathrm{nc}}=0$ and $f(\phi)$ being a symmetric
double well potential. In consequence, many results obtained for the
decomposition of a binary mixture have a counterpart in the dewetting
of thin films and vice versa. The analogy was first noted by Mitlin
\cite{Mitl93} resulting in the notion of 'spinodal dewetting'. Note
that there exist other choices for $f(\phi)$ and $M_{\mathrm{c}}$ in
the Cahn-Hilliard equation \cite{NoSe84,FPT08}.

Equations of similar form may as well model the epitaxial evolution of
surfaces of crystalline solids \cite{SVD91,Savi03,GLSD04,Vved04}.  We
illustrate this employing one of the local models for
Stransk-Krastanov growth found in the literature -- namely a
simplified 'glued wetting-layer model' (for details and derivation see
refs.~\cite[eqs.~18-20]{GLSD04} and \cite{GDN99}). The model assumes
an isotropic wetting energy of the epitaxial film on the solid
substrate, that is added to the (anisotropic) surface energy. Elastic
stresses act through a destabilizing surface stiffness term.  It is
furthermore assumed that a given amount of material is deposited on
the surface that then re-arranges in a process of self-organisation
that might lead to the creation of nano-  or quantum-dots, i.e.,
localized surface structures on the nanometre lengthscale.  In the
case of high-symmetry orientations of a crystal with cubic symmetry
the evolution of the surface profile $\phi(\vec{x},t)$ is described by
Eq.~(\ref{film}) when using small-slope approximation. As the amount
of material is fixed only the conserved part contributes, i.e.,
$M_{\mathrm{c}}>0$ and $M_{\mathrm{nc}}=0$. The model in \cite{GLSD04}
employs a constant mobility $M_{\mathrm{c}}$ (corresponding to a
constant surface diffusion coefficient). Non-constant mobilities might
as well be used. Note that a fourth-order kinematic term is omitted in
the evolution equation as it can lead to artifacts if the slope of the
interface is large (inside the small slope approximation, for details
see \cite{GDN99}).  The free energy functional is
\begin{equation}
F[\phi]\,=\,\int\left[-\frac{\sigma}{2}(\nabla \phi)^2 + 
\frac{\nu}{2}(\Delta \phi)^2 + \frac{a}{12}(\nabla \phi)^4 
+ f(\phi) - \mu\phi\right]dV
\label{eq:en2}
\end{equation}
where $\sigma>0$ is the destabilizing surface stiffness resulting from
elastic stresses, $\nu>0$ represents the energetic cost of corners and
edges, and $a$ quantifies the slope-dependent anisotropic surface
energy (note that we fixed the $b$ of \cite{GLSD04} as $b=a/3$ to
simplify the equation for the present purpose of comparison).  The
local free energy $f(\phi)=\int W_0(\phi)d\phi$ results from the
wetting interaction.  Thereby
$W_0=-w(\phi/\delta)^{-\alpha_w}\,\exp(-\phi/\delta)$ where $\delta$
is a characteristic wetting length, the positive $w$ and $\alpha_w$
characterize the strength and singularity of the underlying
interactions.  The singularity ensures the stability of the stable
monolayer between surface elevations in Stranski-Krastanov growth.
Without wetting interactions the epitaxial film would show
Volmer-Weber growth, i.e., growth would occur in separated islands not
connected by a wetting layer.
Note that we here add the last term to eq.~\ref{eq:en2} where $\mu$ is
the chemical potential.  It does not affect the evolution equation
when $M_{\mathrm{nc}}=0$. Related (in part non-local) equations are employed in
Refs.~\cite{Sieg97,HMRG02,Vved04,PaHu06,PaHu07,TeSp07}.

If vapour deposition is used to deposit the epitaxial layer it is to
expect that equation~(\ref{film}) with (\ref{eq:en2}) and
$M_{\mathrm{nc}}>0$ well describes the process. We are, however, not
aware of such an approach in the literature. Actually, for large
chemical potential $\mu$ as compared to the other terms in $F$
(Eq.~(\ref{eq:en2})) one can even use the system to describe the
evolution under vertical deposition of material. The constant
$M_{\mathrm{nc}}\mu$ does then corresponds to the constant deposition
rate in other models ($V$ in \cite{SVD91}).

Although the overall form of the equation is identical for the various
problems introduced above, the specific physics is very
different. However, still one can employ the same set of techniques to
analyse the various models. Normally, one uses (i) a linear stability
analysis of homogeneous steady states, i.e.\ flat films to determine
the stability of the system and typical length scales that will
dominate the short-time evolution in case the homogeneous state is
unstable. (ii) Depending on the properties of the dispersion relation
obtained in the linear stability analysis one might be able to
analytically study stable and unstable steady state solutions and
their stability in the weakly non-linear regime. This is, however
often not possible.  (iii) In the strongly non-linear regime steady
states and their stability might still be obtained, e.g., using
continuation techniques \cite{DKK91,DKK91b}.  These are readily
available for two-dimensional systems that can be expressed as
ordinary differential equations \cite{AUTO97}.  Recently, they were
also introduced for the full three-dimensional problem, in particular
for Eq.~(\ref{film}) with $M_{\mathrm{nc}}=0$ and (\ref{eq:en1})
\cite{BeTh08}. Note, that variational methods are apt to obtain the
steady states directly from the functional $F$, but are not suitable
to discuss the stability of the steady states as this involves dynamic
aspects.  Many groups prefer to 'skip' step (iii) and rather directly
(iv) simulate the evolution equation in time using advanced numerical
techniques (spectral, pseudo-spectral, or semi-implicite).

We remark here that other types of continuum description exist for all
the mentioned systems. Whereas here we focus on evolution equations
for surface or interface profiles, another class of models describes
interface evolution using phase fields \cite{AMW98}. See, for
instance, for dewetting and liquid films/drops in general
Ref.~\cite{PiPo00} and for epitaxial growth
Ref.~\cite{WLKJ04,HMRG02,Vved04}. We do entirely exclude from our
consideration the vast literature on discrete stochastic models that
exist for dewetting/evaporation processes (e.g.,
\cite{RRGB03,YoRa06,Vanc08} as well as for surface growth (for reviews
see, e.g., \cite{Vics89,Hinr06}).

In the following, we restrict ourselfes to two-dimensional physical
situations described by film thickness profiles $\phi(x,t)$ that
depend on one spatial coordinate only. The drops or quantum-dots in 2d
will actually refer to liquid ridges or quantum-wires in 3d, respectively.
We will perform a basic
analysis of linear stability and steady states 'in parallel' for a
dewetting liquid film on a solid substrate (Section~\ref{sec:dew}), a
dewetting evaporating/condensing liquid film on a solid substrate
(Section~\ref{sec:evap}), and the epitaxial structuring of a solid
film (Section~\ref{sec:epitak}).  Note that we only review steps (i)
and (iii) of the above introduced scheme that sketches a more complete
analysis of the system behaviour.  The next step would be to use
advanced numerical techniques to simulate the evolution of the films
in time for two- and three-dimensional physical settings.  It involves
a rather large number of techniques and groups and we would like to
refer the reader to the individual publications cited in the
respective sections below.

Before we start we would like to point out the relevance of stable and
unstable steady state solutions for systems that evolve in time. Most
steady state solutions are either not linearly stable or do not
correspond to the global energetical minimum that the system will
finally approach. Such solutions are normally not well appreciated in
the literature as they are not present 'in equilibrium'. They are,
however, often present for a long time in the course of the time
evolution and due to their character as saddles in function space 
they do often 'structure' the evolution towards
equilibrium. Their stability properties (i.e., growth and relaxation
rates) determine time-scales for important steps of the dynamics.  and
as important transients in experiments of finite duration they might
actually even get 'frozen in' or 'dried in' as, e.g., in the
dewetting of thin polymer films \cite{Reit92,Seem05} or 
suspensions \cite{TMP98,Mart07}, respectively.
In this connection, coarsening, is a particularly interesting issue as
in its course the system 'passes through' an infinite number of steady
state solutions that are stable when taking their typical size as
reference size, but are unstable with respect to modes on larger
lengthscales, i.e. with respect to coarsening. One could say the
individual solutions do first 'attract a time-evolution' and then
'expel it' along the single unstable direction (corresponding to
coarsening). 

\section{Dewetting}
\mylab{sec:dew}

A dewetting film of non-volatile liquid on a solid substrate is
modelled by Eqs.~(\ref{film}) and (\ref{eq:en1}) with
$M_{\mathrm{nc}}=0$ \cite{Mitl93,ODB97,KaTh07}. The local energy
$f(\phi)$ corresponds to a disjoining pressure
$\Pi(\phi)=-\partial_\phi f(\phi)$ \cite{DLP60,deGe85,Isra92}. Various
functional forms are used for $f(\phi)$. The particular choice is,
however, not very relevant for the qualitative behaviour of the
system. The latter only depends on the number and relative
depth/height of the extrema of $f(\phi) - \mu\phi$. We choose here
$f(\phi)=-\kappa[1/(2\phi^2)-b^3/(5\phi^5)]$ as derived by Pismen from
a modified Lennard-Jones potential with hard-core repulsion
\cite{Pism01,PiPo04,PiTh06}. The resulting $f(\phi)$ has only one
minimum at a finite $\phi_{\mathrm{precursor}}=b$. The parameter
$\kappa$ corresponds to a typical energy scale. With the chosen signs
the first term corresponds to a destabilizing long-range van der Waals
interaction (i.e., $\kappa$ is proportional to a Hamacker constant)
whereas the second one represents a short-range stabilizing
interaction. In consequence, the model may describe drops of a
partially wetting liquid in co-existence with a precursor film of
thickness $b$. In the following we scale $\phi$ by the precursor film
thickness, i.e., we fix $b=1$.

If there existed a second minimum at larger finite thickness it would
correspond to a critical height for the transition between
spherical-cap-like drops and pancake-like drops (e.g., the
capillary length when gravity is included) \cite{deGe85,Thie01}. For a
selection of other pressure terms see,
e.g.~\cite{deGe85,TDS88,PiPo00,OrRo92,Mitl93,Shar93b,ShRe96,TMP98,TNPV02,Seem05}.

Inspecting Eq.~(\ref{film}) with $M_{\mathrm{nc}}=0$ one notes that
any flat film (thickness $\phi=\phi_0$) corresponds to a steady state
solution of the system. However, those films might not be stable:
We linearize the system about the flat film employing
harmonic modes, i.e., $\phi(x,t)=\phi_0+\epsilon\,\exp(\beta t + i
kx)$ where $\epsilon\ll1$ is the smallness parameter, $\beta$ the
growth rate of the harmonic mode of wavenumber $k$. Entering this
ansatz in Eq.~(\ref{film}) with (\ref{eq:en1}) gives the dispersion
relation
\begin{equation}
\beta(k)\,=\,-M_{\mathrm{c}}^0\,\gamma\,k^2\,(k^2-k_c^2),
\label{eq:disp1}
\end{equation}
where $M_{\mathrm{c}}^0=M_{\mathrm{c}}(\phi=\phi_0)$. The critical
wavenumber is given by $k_c=\sqrt{-\partial_{\phi\phi}f|_{\phi=\phi_0}/\gamma}$.
For $\partial_{\phi\phi}f|_{\phi=\phi_0}<0$ the film is linearly unstable
for $0<k<k_c$. The most dangerous instability mode, i.e., fastest growing
mode has $k_{\mathrm{max}}=k_c/\sqrt{2}$ and
$\beta_{\mathrm{max}}=M_{\mathrm{c}}^0\gamma k_c^4/4$.  The onset of the
instability occurs at $\partial_{\phi\phi}f|_{\phi=\phi_0}=0$ with
$k_c^{\mathrm{onset}}=0$, i.e., it is a long-wave instability.
Note that mass conservation implies $\beta(k=0)=0$.
Examples of dispersion relations above and below the instability
treshhold are given in Fig.~\ref{fig:dew-disp}.

\begin{figure}
\includegraphics[width=0.8\hsize]{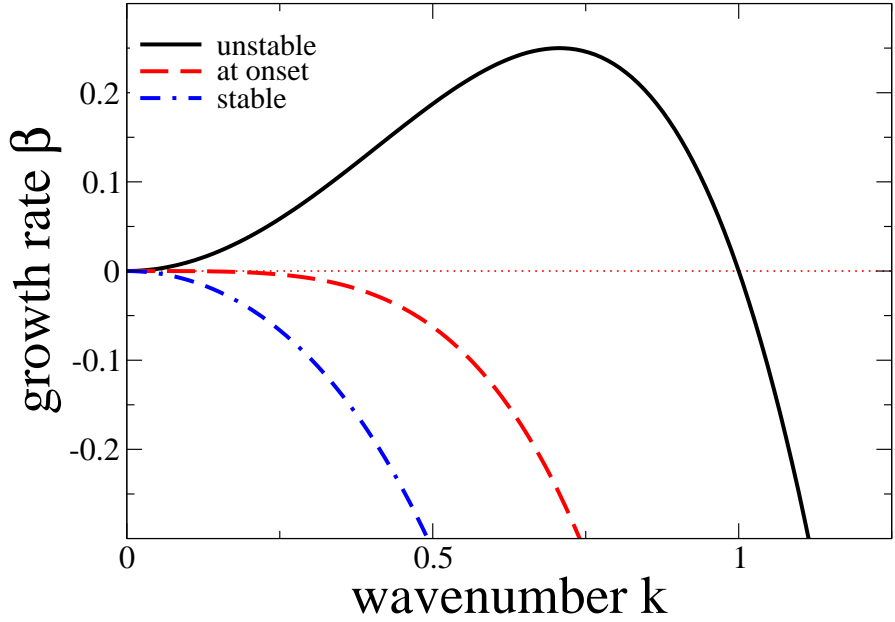}
\caption{Dispersion relations for the instability of a flat liquid film
w.r.t.\ surface modulations resulting in dewetting and the evolution of patterns 
of droplets. Shown is a stable and an unstable case in the generic form
$\beta=-k^2(k^2-k_c^2)$, i.e., the growth rate $\beta$ is scaled by $M_{\mathrm{c}}^0\,\gamma$.
The wave number is given in units of $k_c$.
}
\mylab{fig:dew-disp}
\end{figure}

\begin{figure}
\includegraphics[width=0.45\hsize]{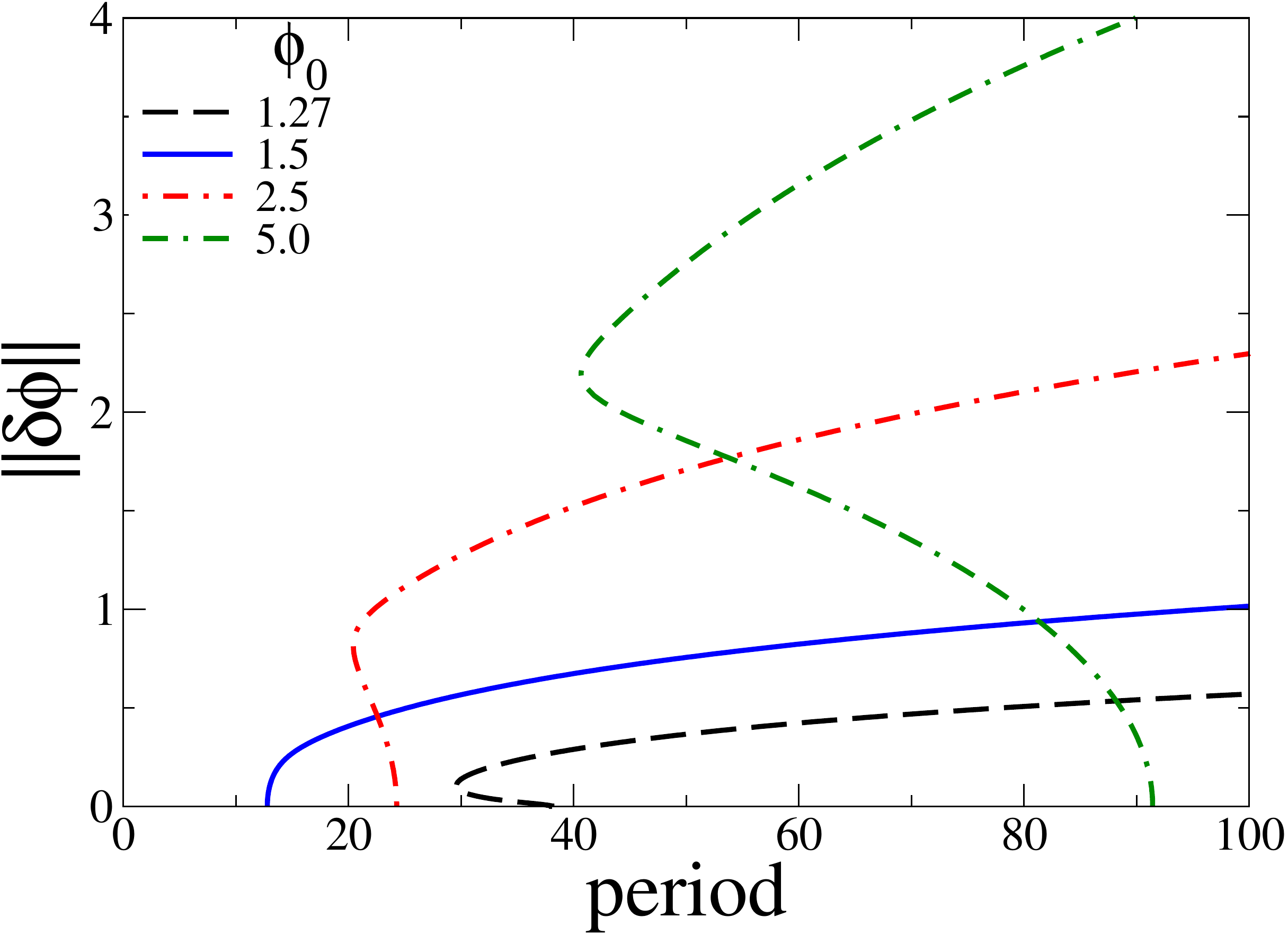}\hspace{0.05\hsize}
\includegraphics[width=0.45\hsize]{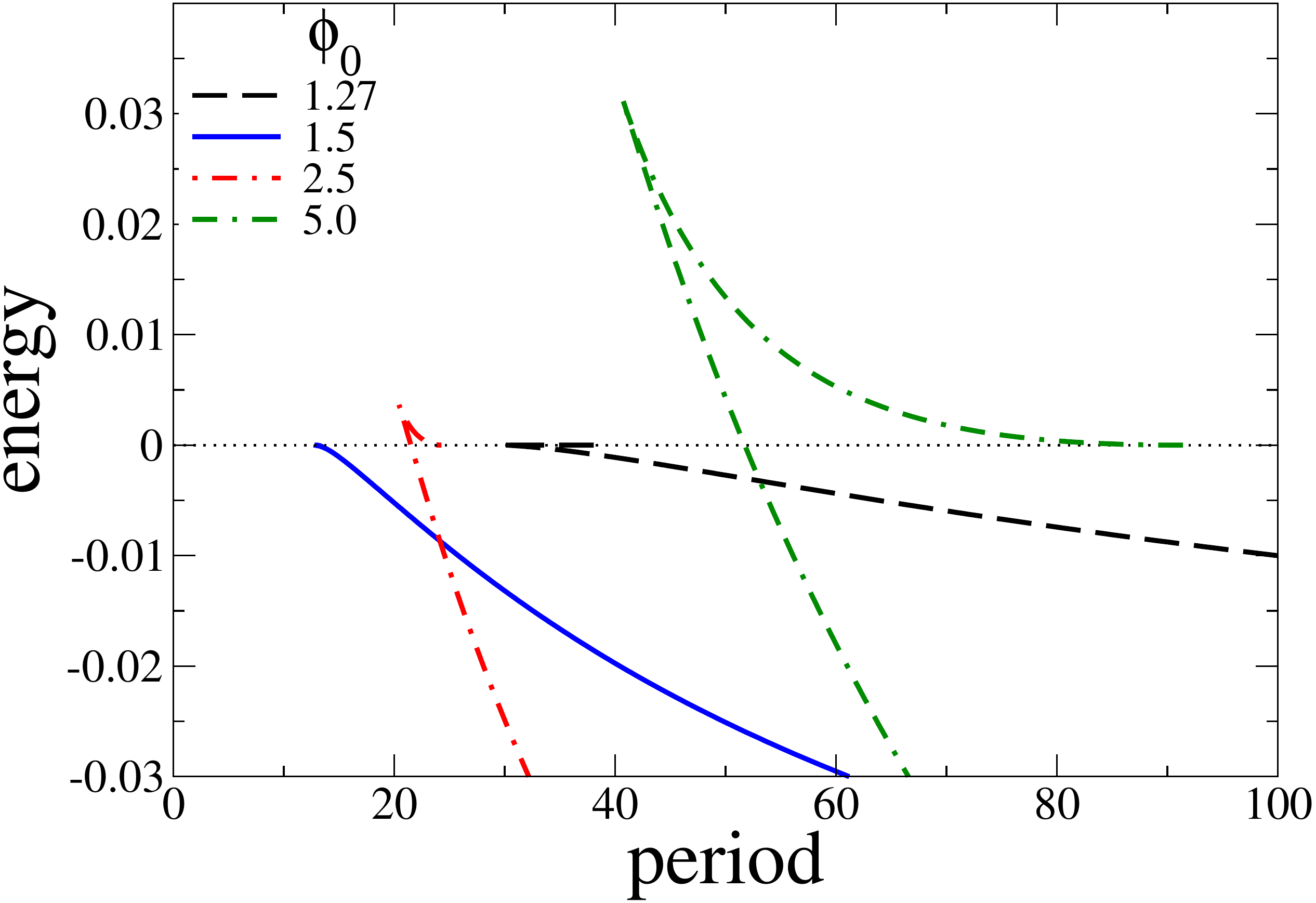}
\caption{Shown are characteristics of families of steady
  one-dimensional droplet structures arising in dewetting of a
  non-volatile liquid using a simple disjoining pressure. The left
  panel gives the $L_2$-norm $||\delta\phi||$ and the right one the
  energy $F$ (Eq.~(\ref{eq:en1})) per length.  The legends give the
  corresponding mean film thicknesses.  The families are obtained
  using continuation techniques. Note that the solutions are unstable
  w.r.t.\ coarsening.}
\mylab{fig:dew-fam}
\end{figure}

\begin{figure}
\includegraphics[width=0.8\hsize]{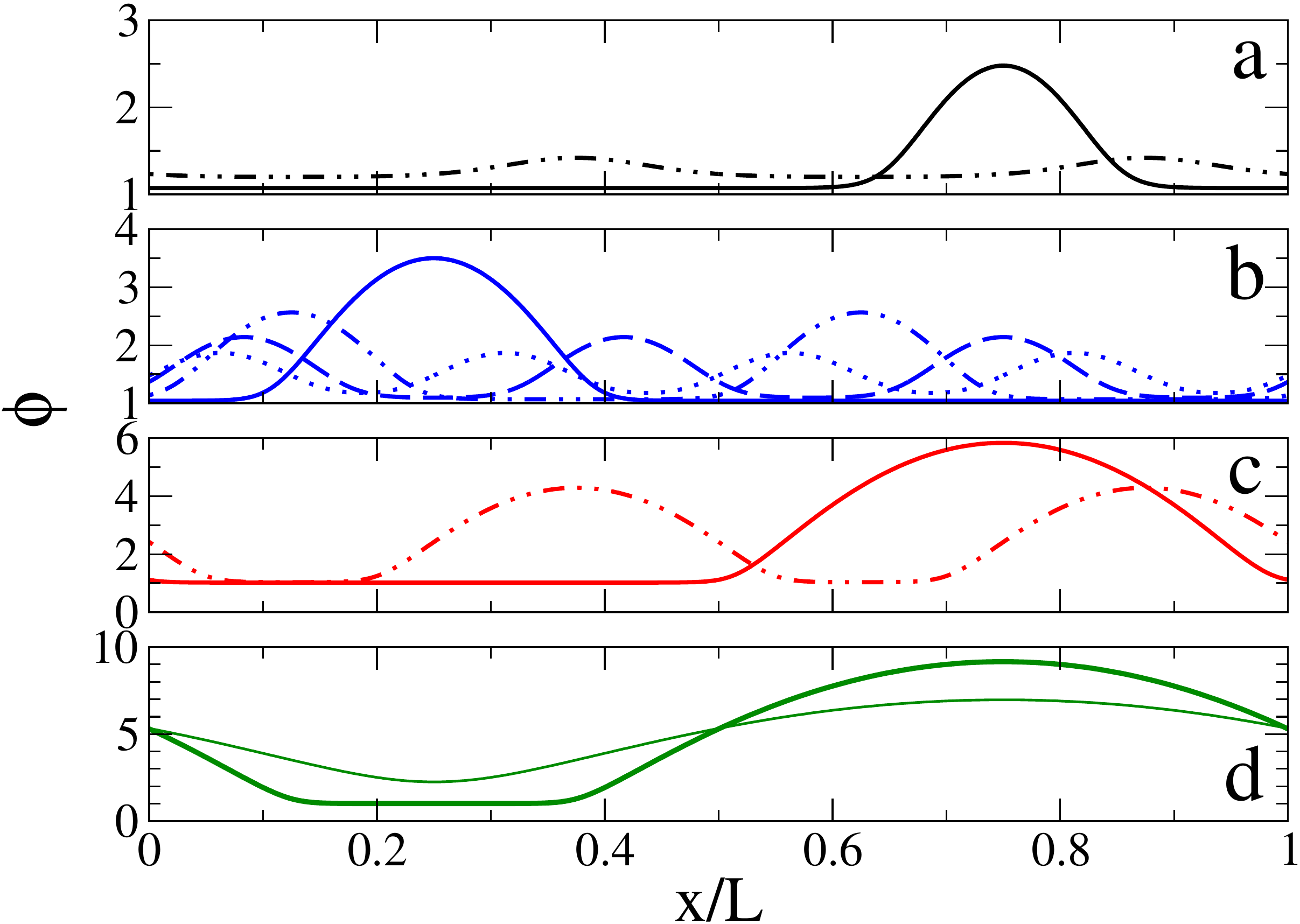}
\caption{Shown are examples of film thickness profiles for the various
  branches presented in Fig.~\ref{fig:dew-fam}. Mean thicknesses
  $\bar{\phi}$ are (a) 1.27, (b) 1.5, (c) 2.5, and (d) 5.0.  The system
  size is $L=60$ in all cases. Given are profiles from the following
  branches: $n=1$ (solid lines), $n=2$ (dot-dashed lines), $n=3$
  (dashed line), and $n=4$ (dotted line).  The thin line in (d)
  corresponds to a nucleation solution. For given $L$ only the $n=1$
  profiles are stable w.r.t.~coarsening.}
\mylab{fig:dew-prof}
\end{figure}

Steady film thickness profiles are obtained by solving
Eq.~(\ref{film}) with $\partial_t \phi=0$.  Note that in the present
setting this corresponds to $\delta F / \delta \phi = 0$ because the
first integration constant (when integrating Eq.~\ref{film}) is zero
as we look at systems without large-scale meanflow.
The whole solution structure might be mapped out in parameter space
employing continuation techniques using analytically or numerically
known solutions to start the continuation
\cite{DKK91,DKK91b,AUTO97,KaTh07}. Many parameters might be used as
main continuation parameter, e.g., period or domain size, film
thickness, or any parameter of the energy. Here we only use system
size. For a more detailed explanation of the continuation procedure
for thin film equations see the Appendices of \cite{TBBB03,JBT05}.

For a linearly unstable film a one parameter family of periodic film
profiles bifurcates from the flat film at wavenumber $k_c$ as
illustrated in Fig.~\ref {fig:dew-fam}. At the bifurcation point a
steady harmonically modulated surface profile exists with a period of
$2\pi/k_c$ and an infinitesimally small amplitude. Using period as
main continuation parameter the whole family of periodic solutions can be
obtained. A selection of resulting film thickness profiles is given in
Fig.~\ref{fig:dew-prof}.  We characterize solutions by their norm
$||\delta \phi||=[(1/L)\int_0^L (\phi(x)-\phi_0)^2dx]^{1/2}$ and
energy per length $E=(1/L)\int_0^L F[\phi]dx$.
Note that mass conservation advises us to directly compare only
profiles of identical mean film thickness. This is ensured by $\mu$
acting as a Lagrange multiplier. In consequence, $\mu$ changes along
the individual branches in such a way that the mean of $\phi(x)$ is
kept fixed. 

The resulting branches might bifurcate from the trivial solution
supercritically, i.e., they bifurcate towards the region of the
linearly unstable flat film (forward towards larger periods; see,
e.g., in Fig.~\ref{fig:dew-fam} curve for $\phi_0=1.5$).  Or the
branch bifurcates subcritically, i.e., it emerges towards the region
of the linearly stable flat film (backward towards smaller periods;
see, e.g., in Fig.~\ref{fig:dew-fam} curves for $\phi_0=1.27, 2.5$, or
$5.0$). The location of the border between sub- and supercritical
behaviour can be determined employing a weakly nonlinear analysis and
can be expressed as an algebraic condition for the 2nd, 3rd and 4th
derivative of $f(\phi)$ \cite{TVK06}.

In the case of the subcritical bifurcation, the subcritical branch
(i.e., between the bifurcation and the saddle-node bifurcation where
it 'folds back' towards larger periods) consists of unstable,
nucleation solutions that aquire an importance for the rupture process
of the film if the system is noisy or 'dirty': Note that for the
chosen potential $f(\phi$) there exist no metastable films. However,
one may distinguish two thickness ranges within the linearly unstable
range: (i) the linearly unstable modes are fast and will dominate the
time evolution even in the presence of defects (finite size
perturbations); (ii) the linear modes are slow, and defects -- if there
are any present -- will dominate the time evolution. The
distinction is related to the existence of the subcritical branch of
unstable solutions. They correspond to nucleation or threshold
solutions as they have to be overcome to break the film into drops
smaller than the critical wavelength of the linear instability.  As
they are saddles in function space they can 'organize' the evolution
of defect-ridden thin films by offering a fast track to film
rupture. This allows to determine a typical time for nucleation events
even inside the linear unstable regime and finally to distinguish the
nucleation-dominated and instability-dominated behaviour of linearly
unstable thin films \cite{TVN01,TVNP01}. A detailed account and
comparison to the results of \cite{Beck03} is found in
Ref.~\cite{Thie03}. Note that similar ideas have since been applied to
the break-up of liquid ribbons \cite{DiKo07}. Recently we also
performed a more detailed analysis for a three-dimensional systems
\cite{BeTh08,BHT09}.

For unstable flat films of thicknesses $\phi_0$ the respective
branches of solutions bifurcating at $L_{c}=2\pi/k_c$ shown in
Fig.\,\ref{fig:dew-fam} are only the first of an respective infinite
number of primary solution branches. These bifurcate at domain sizes
$L_{cn}=2\pi n/k_c$, $n=1,2,\dots$.  The branch bifurcating at
$L_{cn}$ consist of the $n=1$ branch 'stretched in $L$' by a factor
$n$. The actual thickness profiles of the $n$ branch consist of $n$
identical drops. This representation of the periodic solutions in
dependence of domain size might seem pointless for the present
problem.  The reason is that the different branches are entirely
decoupled. As the situation changes strongly when either looking at
other energy functionals (see below Section~\ref{sec:epitak}), or when
breaking the reflection or translational symmetry of the system (by
including a lateral driving force \cite{GNDZ01,Thie01,ThKn04},
substrate heterogeneity \cite{TBBB03}, or lateral boundary conditions
\cite{TMF07}) it is, however, useful to introduce the various branches
here.

The monotonous decrease of energy (on the low energy branch) with
system size indicates that the system tends to coarse towards larger
and larger structure size. The steady solutions on the 'stable'
branches (the branches of drop solutions that continue towards
infinite period) are stable when looked at in a domain of the size of
their period, but unstable on larger domains, i.e., they are saddles
in function space that form the 'stepping stones' of the coarsening
process: They first 'attract the time-evolution' and then 'expel it'
along the only unstable direction (corresponding to coarsening).  The
coarsening behaviour for a dewetting film is discussed in more detail
in Refs.~\cite{BGW01,BKTB02,GlWi03,GlWi05}.  Related results for the
Cahn-Hilliard equation are discussed, e.g., in
Refs.~\cite{Lang92,ToPo02,KoOt02}.

Note that results on the nonlinear stability of flat films and the
related steady state solutions on branches not connected to the
trivial flat film solutions are not discussed here (but see
\cite{Thie03,ThKn04,KaTh07}). For discussions of the evolution in time
of dewetting films in two- and three-dimensional settings we refer the
reader to
Refs.~\cite{ODB97,ShKh98,Oron00,BeNe01,Thie02,BPT03,Beck03,VSKB05,DiKo07,BeTh08}.
Note finally that continuation may not only be used to obtain families
of steady or stationary profiles
\cite{TNPV02,Thie01,JBT05,TVK06,PTTK07b}, but as well to track their
stability and bifurcations \cite{ThKn03,ThKn04,Sche05,ThKn06}.

\section{Evaporation}
\mylab{sec:evap}

When including evaporation in the presently studied framework one uses
Eq.~(\ref{film}) with the energy functional~(\ref{eq:en1}) and
$M_{\mathrm{nc}}\neq 0$.  One notes that unlike the case without
evaporation flat films of arbitrary thickness $\phi=\phi_0$ do not
correspond to a steady state solution of the system any more.
However, steady flat film solutions still exist when wettability and
evaporation balance. For a given chemical potential the corresponding
steady thicknesses are given by
\begin{equation}
\partial_\phi f|_{\phi=\phi_0} = \mu,
\label{eq:film_evap}
\end{equation}
i.e., for the disjoining pressure used here
$\phi_0=[\kappa(1\pm(1-4b^3\mu/\kappa))^{1/2}/2\mu]^{1/3}$. For
condensation ($\mu>0$) two such equilibria exist if $\mu<\kappa/4b^3$
whereas for evaporation ($\mu<0$) only one thickness exists. Note that
this might be different for qualitatively different disjoining pressures like
the one used in \cite{LGP02}. As above the steady films might be unstable.

We linearize Eq.~(\ref{film}) with (\ref{eq:en1})
about the flat film of thickness $\phi_0$  given by Eq.~(\ref{eq:film_evap}) employing 
harmonic modes as above to obtain the dispersion relation
\begin{equation}
\beta(k)\,=\,-M_{\mathrm{c}}^0\,\gamma\left(k^2+
\frac{M_{\mathrm{nc}}^0}{M_{\mathrm{c}}^0}\right)\,(k^2-k_c^2)
\label{eq:disp2}
\end{equation}
where $M_{\mathrm{nc}}^0=M_{\mathrm{nc}}(\phi=\phi_0)$,
$M_{\mathrm{c}}^0=M_{\mathrm{c}}(\phi=\phi_0)$, and
$k_c=\sqrt{-\partial_{\phi\phi}f|_{\phi=\phi_0}/\gamma}$ as without
evaporation.  As the expression in the first parenthesis is always
positive the film is linearly unstable for
$\partial_{\phi\phi}f|_{\phi=\phi_0}<0$ in the wave number range
$0<k<k_c$. The most dangerous instability mode occurs for
$k_{\mathrm{max}}=\sqrt{(k_c^2-M_{\mathrm{nc}}^0/M_{\mathrm{c}}^0)/2}$
with
$\beta_{\mathrm{max}}=M_{\mathrm{c}}^0\,\gamma(M_{\mathrm{nc}}^0/M_{\mathrm{c}}^0+k_c^2)^2/4$
if the expression under the square root is positive. Otherwise the
mode with $k=0$ grows fastest ($\beta(k=0)=M_{\mathrm{nc}}^0k_c^2$).

The onset of the instability occurs at
$\partial_{\phi\phi}f|_{\phi=\phi_0}=0$ with $k_c^{\mathrm{onset}}=0$, i.e.,
it is a long-wave instability.  Qualitatively different examples of
dispersion relations above and below the instability treshold are
given in Fig.~\ref{fig:evap-disp}.

\begin{figure}
\includegraphics[width=0.8\hsize]{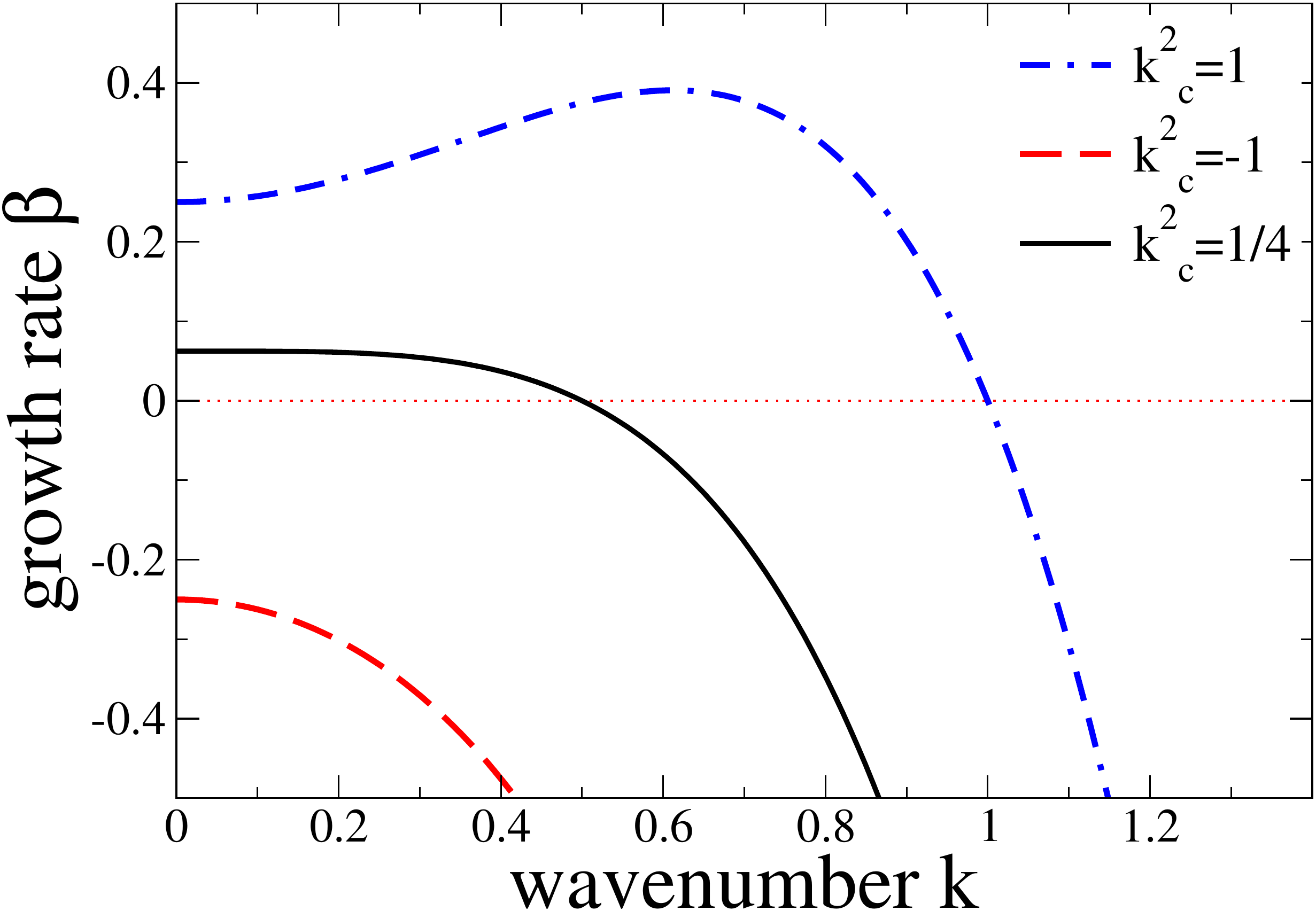}
\caption{Dispersion relations for the instability of a flat volatile
  liquid film w.r.t.~surface modulations resulting in the evolution of
  a pattern of droplets. Shown are three qualitatively different cases in
  the generic form $\beta=-(k^2+M_{\mathrm{nc}}^0/M_{\mathrm{c}}^0)\,(k^2-k_{c}^2)$, i.e., we fix
$\gamma M_{\mathrm{c}}^0=1$. We furthermore choose $M_{\mathrm{nc}}^0/M_{\mathrm{c}}^0=1/4$.}
\mylab{fig:evap-disp}
\end{figure}

\begin{figure}
\includegraphics[width=0.45\hsize]{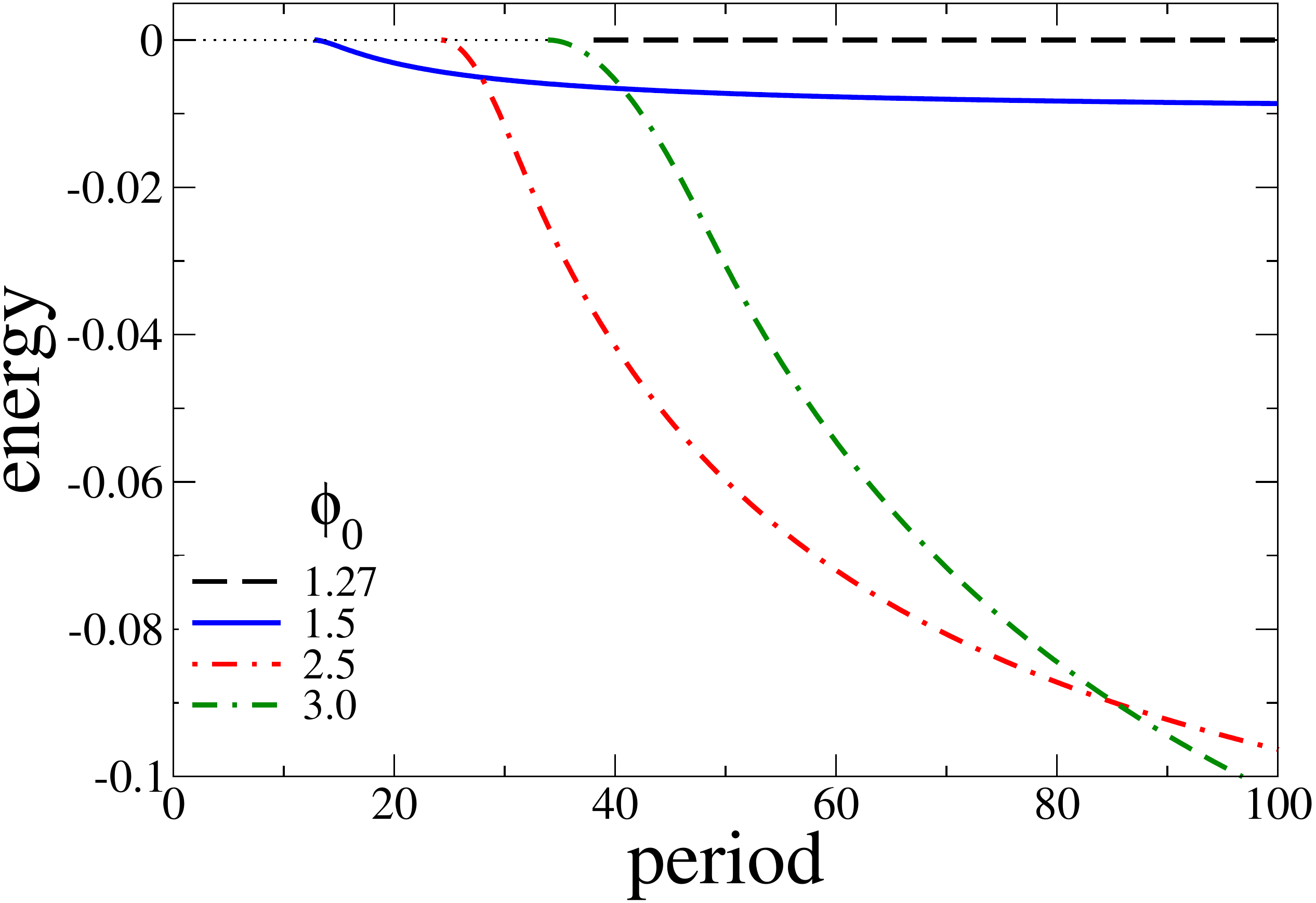}\hspace{0.05\hsize}
\includegraphics[width=0.45\hsize]{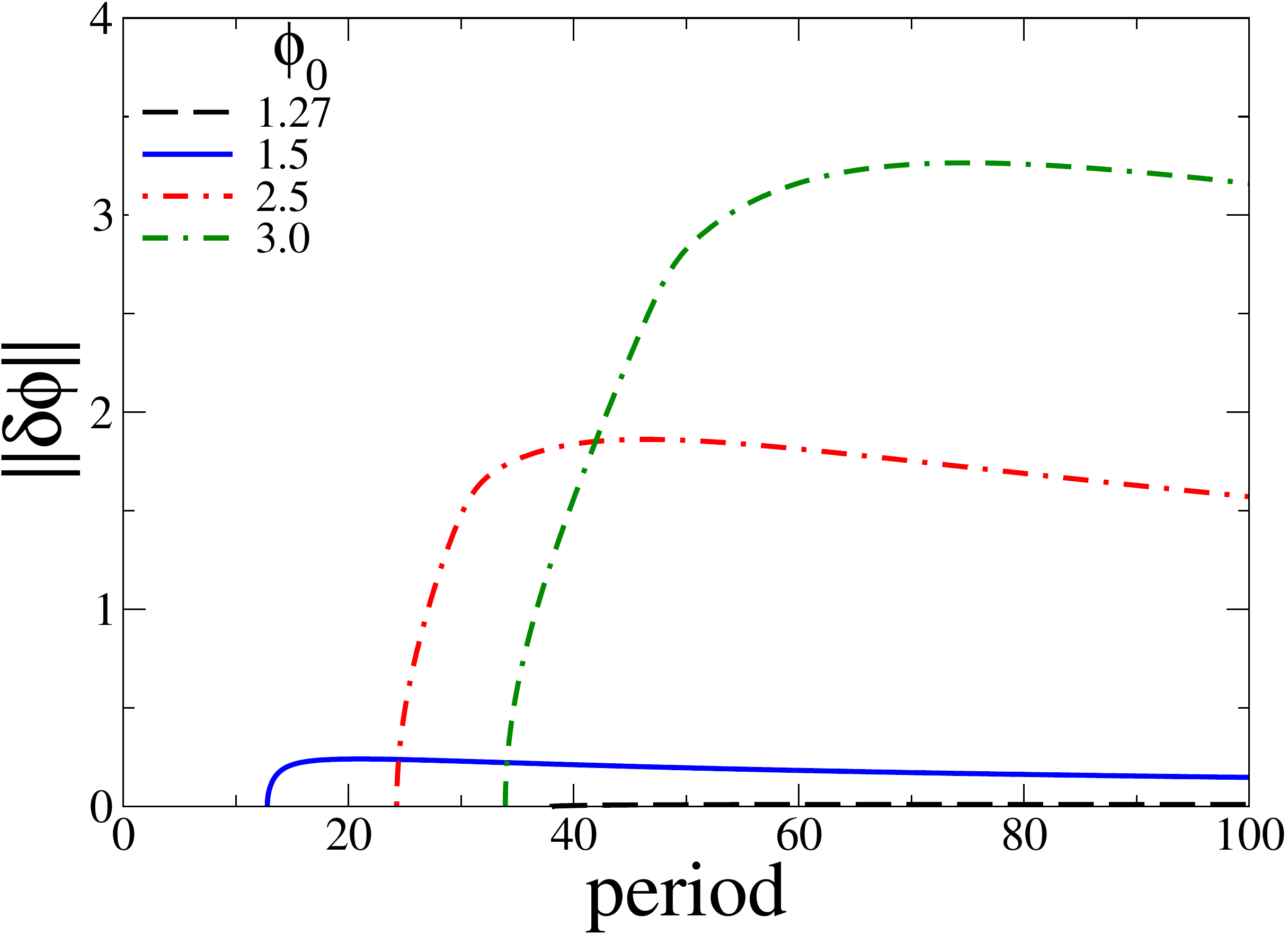}\\
\includegraphics[width=0.45\hsize]{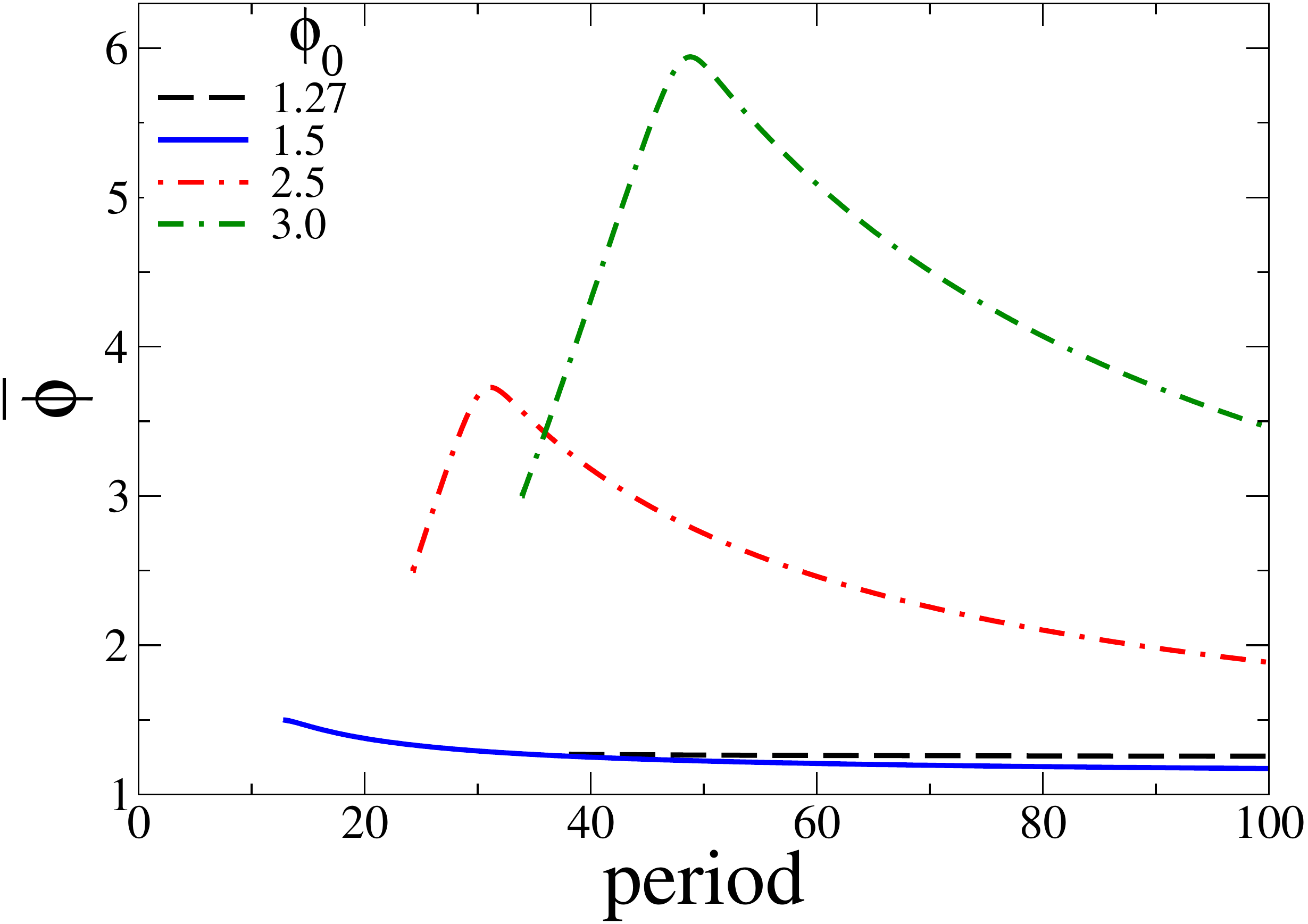}\hspace{0.05\hsize}
\includegraphics[width=0.45\hsize]{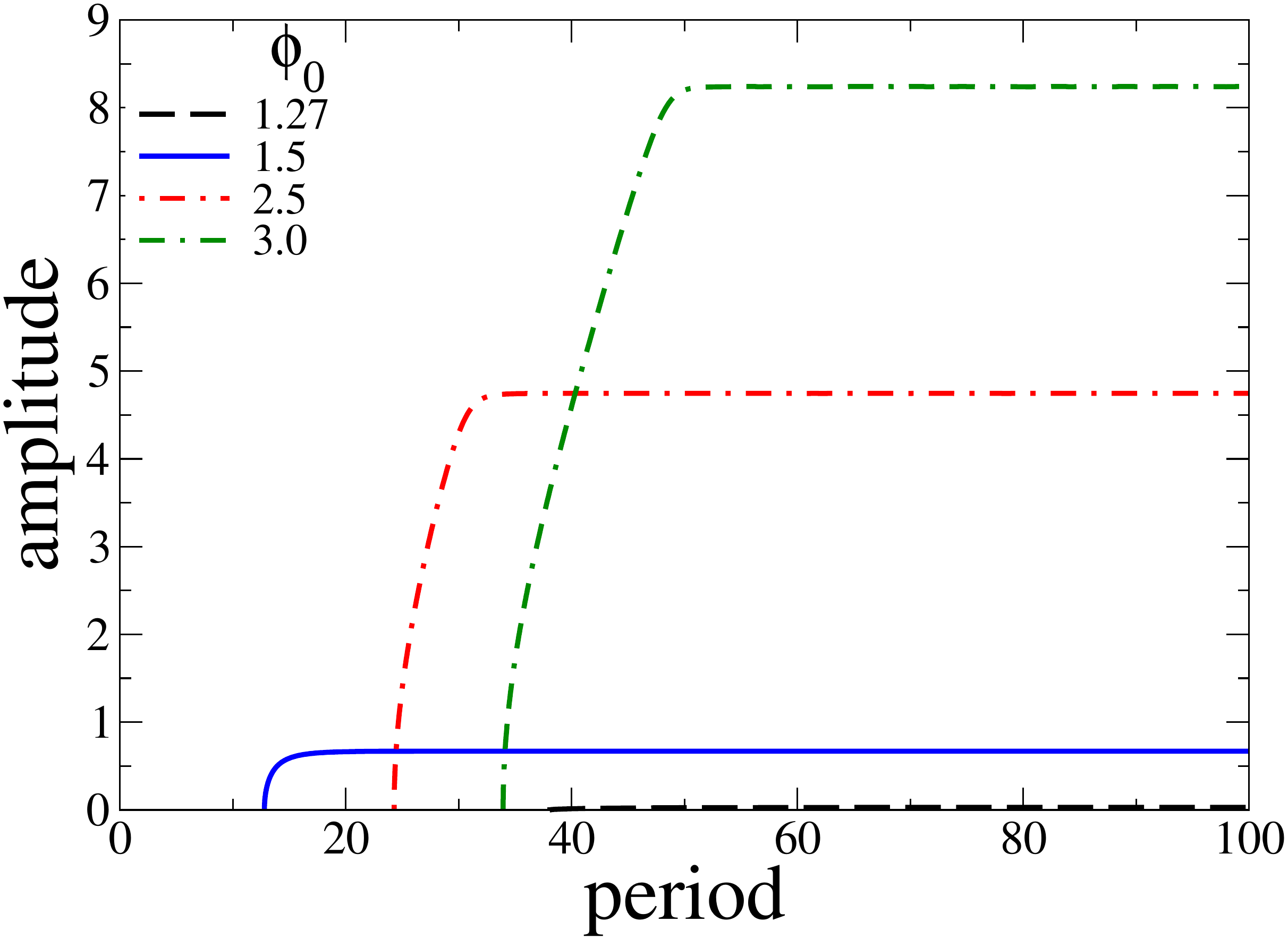}
\caption{Shown are characteristics of families of steady
  one-dimensional droplet structures arising in dewetting of a
  volatile liquid using a simple disjoining pressure. From top left to
  bottom right the panels give the $L_2$-norm, the energy $F$
  (Eq.~(\ref{eq:en1})) per length, the mean film thickness $\bar{\phi}$,
  and the droplet amplitude.  The legends give the corresponding mean
  film thicknesses for the flat film $\phi_0$ at the identical
  chmicalpotential $\mu$.  The families are obtained using
  continuation techniques. Note that the solutions are unstable
  w.r.t.\ coarsening.}
\mylab{fig:evap-fam}
\end{figure}

\begin{figure}
\includegraphics[width=0.8\hsize]{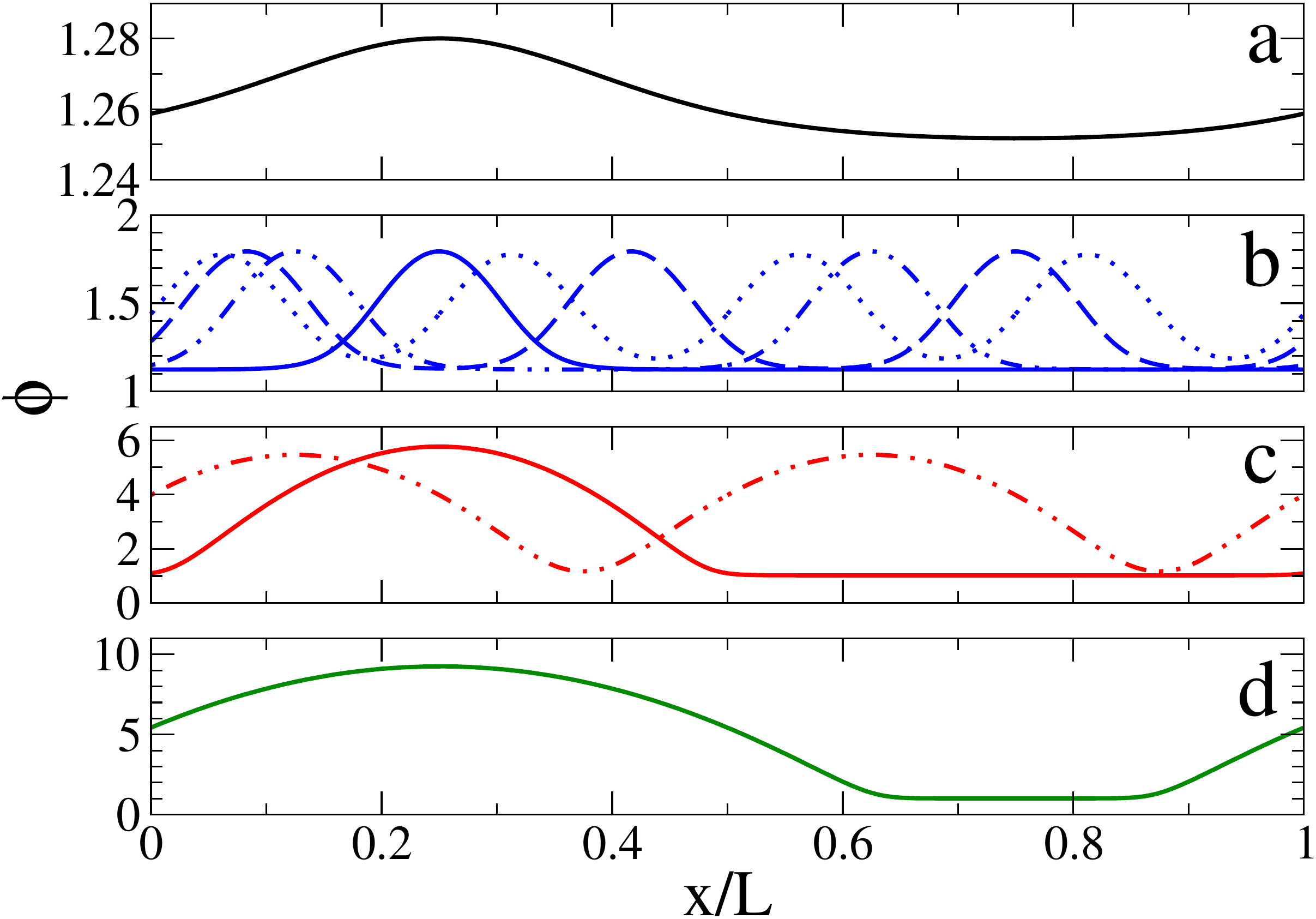}
\caption{Shown are examples of film thickness profiles for the various
  branches presented in Fig.~\ref{fig:evap-fam}. Mean thicknesses
  $\bar{\phi}$ are (a) 1.27, (b) 1.5, (c) 2.5, and (d) 3.0.
  The system size is $L=60$ in all cases. Given are profiles
  from the following branches: $n=1$ (solid lines), $n=2$ (dot-dashed
  lines), $n=3$ (dashed line), and $n=4$ (dotted line).}
\mylab{fig:evap-prof}
\end{figure}

As above steady thickness profiles are obtained by setting $\partial_t
\phi=0$ in Eq.~(\ref{film}). For a linearly unstable film a one
parameter family of profiles bifurcates from the flat film at $k_c$ as
illustrated in the four panels of Fig.~\ref {fig:evap-fam}. In
contrast to the case without evaporation/condensation here the mean
film thickness changes along the branches whereas the imposed chemical
potential remains constant. A selection of film thickness
profiles is given in Fig.~\ref{fig:evap-prof}.

Most notably the resulting branches always bifurcate supercritically
from the trivial solution. The results furthermore indicate that there
is no metastable film thickness range at all, implying that nucleation
does not play any role for a condensing/evaporating film described by
the present disjoining pressure. This, however, needs a deeper
analysis in the future including a comparison of the behaviour for
different pressure terms.
Remarkably, the amplitude (droplet height) approaches a limiting value
for a domain size of about $3L_c/2$. As the energy monotonously decreases
with domain size we expect these solutions to be unstable
w.r.t.~coarsening as the droplet pattern in the non-volatile case.
Note finally, that everything discussed above for non-volatile films 
regarding the infinite number of primary solution branches as well applies
to the branches shown in Fig.~\ref{fig:evap-fam}.

After the short analysis of the the volatile and non-volatile thin
liquid film system we next focus on a thin film equation that
describes the evolution of a solid film.

\section{Epitaxial growth}
\mylab{sec:epitak}

The final example is a thin film model for the epitaxial evolution of
surfaces of crystalline solids for a fixed ammount of deposited
material that is illustrated here employing a simplified 'glued
wetting-layer model' \cite{GLSD04}.  The evolution
equation~(\ref{film}) describes such a system when combined with the
energy functional~(\ref{eq:en2}) and $M_{\mathrm{nc}}= 0$.  As in the
case of the non-volatile liquid film material is conserved and any
flat film of thickness $\phi=\phi_0$ corresponds to a steady state
solution of the system.

As those trivial steady states might be unstable we perform a linear
stability analysis along the lines of the previous sections and obtain
the dispersion relation
\begin{equation}
\beta(k)\,=\,-M_{\mathrm{c}}^0\nu\,\,k^2\left(k^2-k_{c1}^2\right)\,(k^2-k_{c2}^2)
\label{eq:disp3}
\end{equation}
with 
\begin{equation}
k_{c1/c2}=\sqrt{\frac{\sigma}{2\nu}\left(1\pm
\sqrt{1-\frac{4\nu\,\partial_{\phi\phi}f|_{\phi=\phi_0}}{\sigma^2}}\right)}
\label{eq:kc3}
\end{equation}

The film is linearly unstable for
$\partial_{\phi\phi}f|_{\phi=\phi_0}<\sigma^2/4\nu$ in the wave number
range $k_{c1}<k<k_{c2}$. The fastest growing mode has the wavenumber
$k_{\mathrm{max}}=(\sigma/3\nu)^{1/2}\sqrt{1+\sqrt{1-3\nu\partial_{\phi\phi}f|_{\phi=\phi_0}/\sigma^2}}$.
For values of $\partial_{\phi\phi}f|_{\phi=\phi_0}<0$ one of the two
critical wavenumbers becomes imaginary, i.e., the film is then
unstable for $0<k<k_{c2}$.  Note that the onset of the instability
occurs at $\partial_{\phi\phi}f|_{\phi=\phi_0}=\sigma^2/4\nu$ with
$k_c^{\mathrm{onset}}=\sqrt{\sigma/2\nu}$, i.e., it is a short-wave
instability.
Qualitatively different examples of actual dispersion curves
above and below the instability treshold are given in
Fig.~\ref{fig:epi-disp}.

\begin{figure}
\includegraphics[width=0.8\hsize]{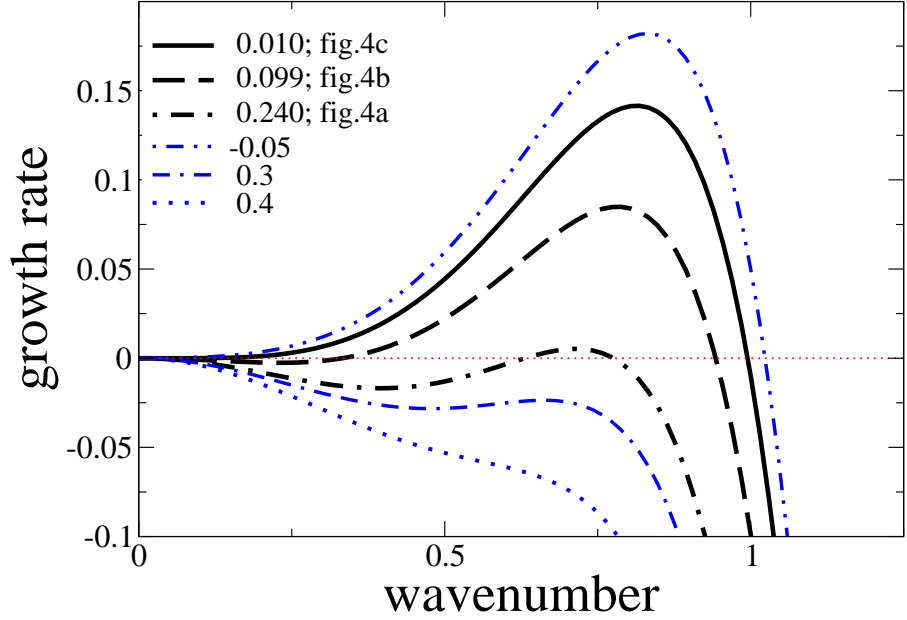}
\caption{Dispersion relations for the instability of a flat epitaxial
  film w.r.t.~surface modulations resulting in the growth of quantum
  dots.  Used is the simplest model proposed in
  Ref.~\cite{GLSD04}. The legend gives the dimensionless value of
  $\partial_\phi f$ at $\phi_0$. Starting with the lowest one the three
  bold curves correspond to the cases studied in Figs.~4(a), (b) and
  (c) of \cite{GLSD04}. }
\mylab{fig:epi-disp}
\end{figure}

\begin{figure*}
\includegraphics[width=0.45\hsize]{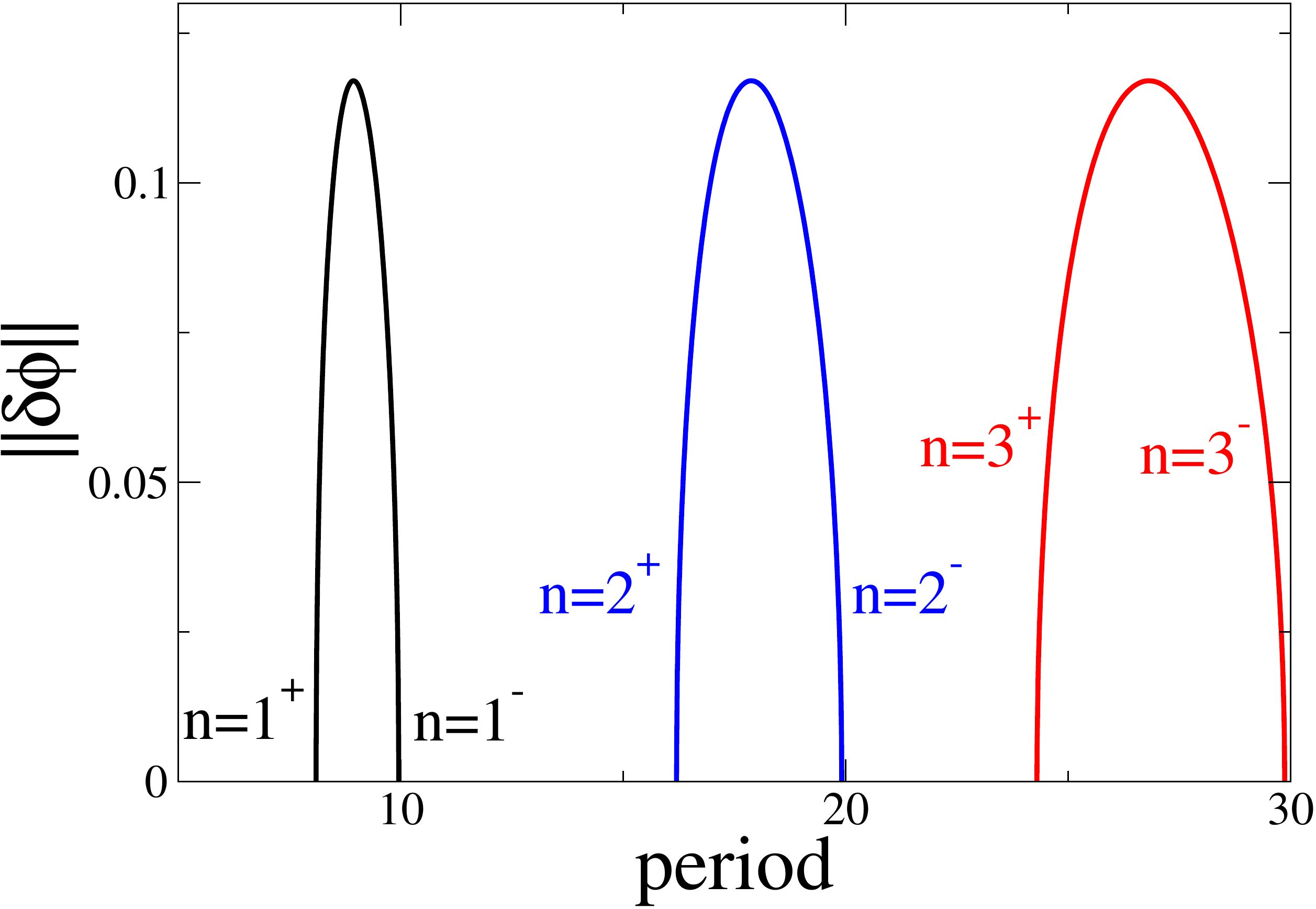}\hspace{0.05\hsize}
\includegraphics[width=0.45\hsize]{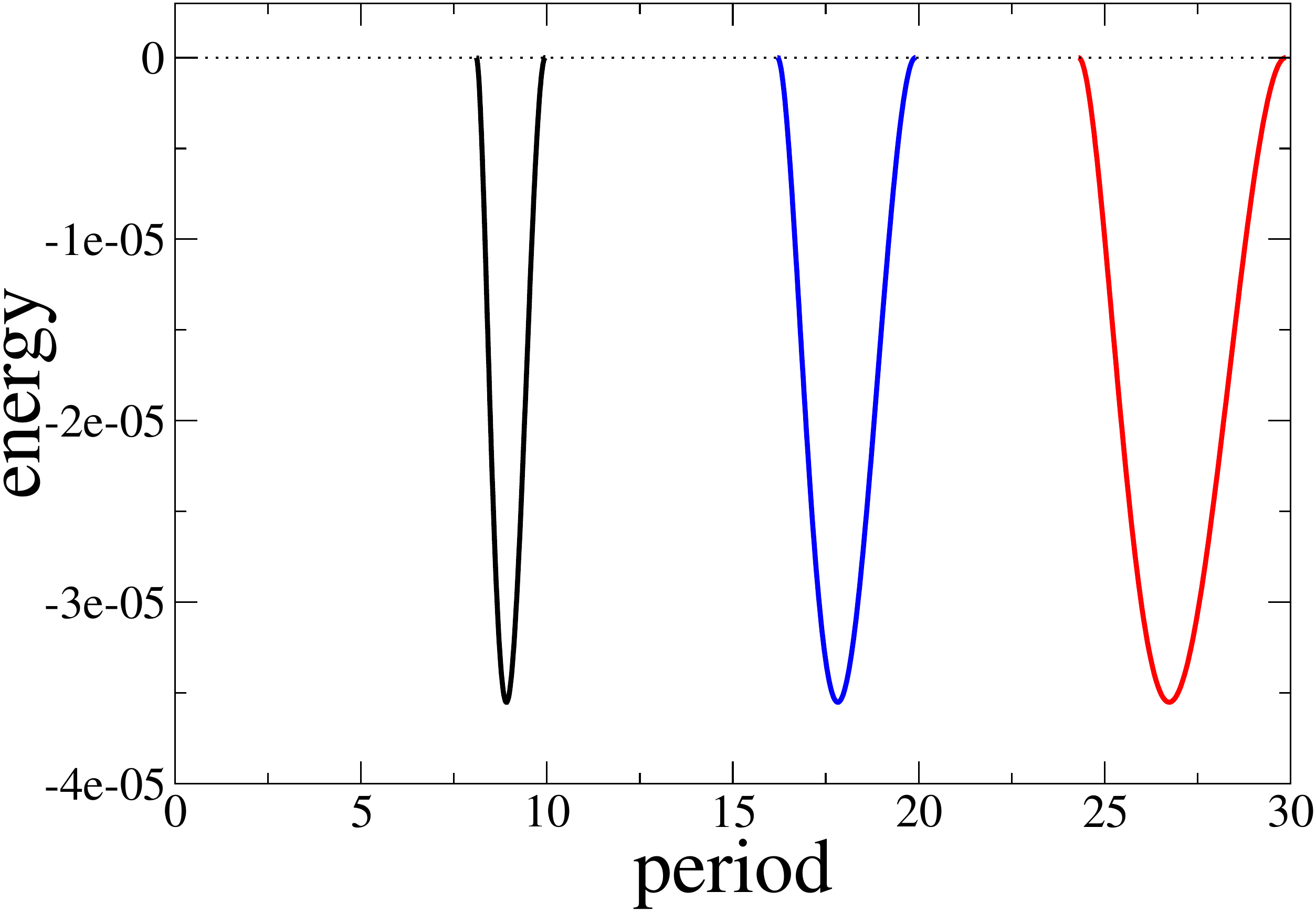}\\
\includegraphics[width=0.45\hsize]{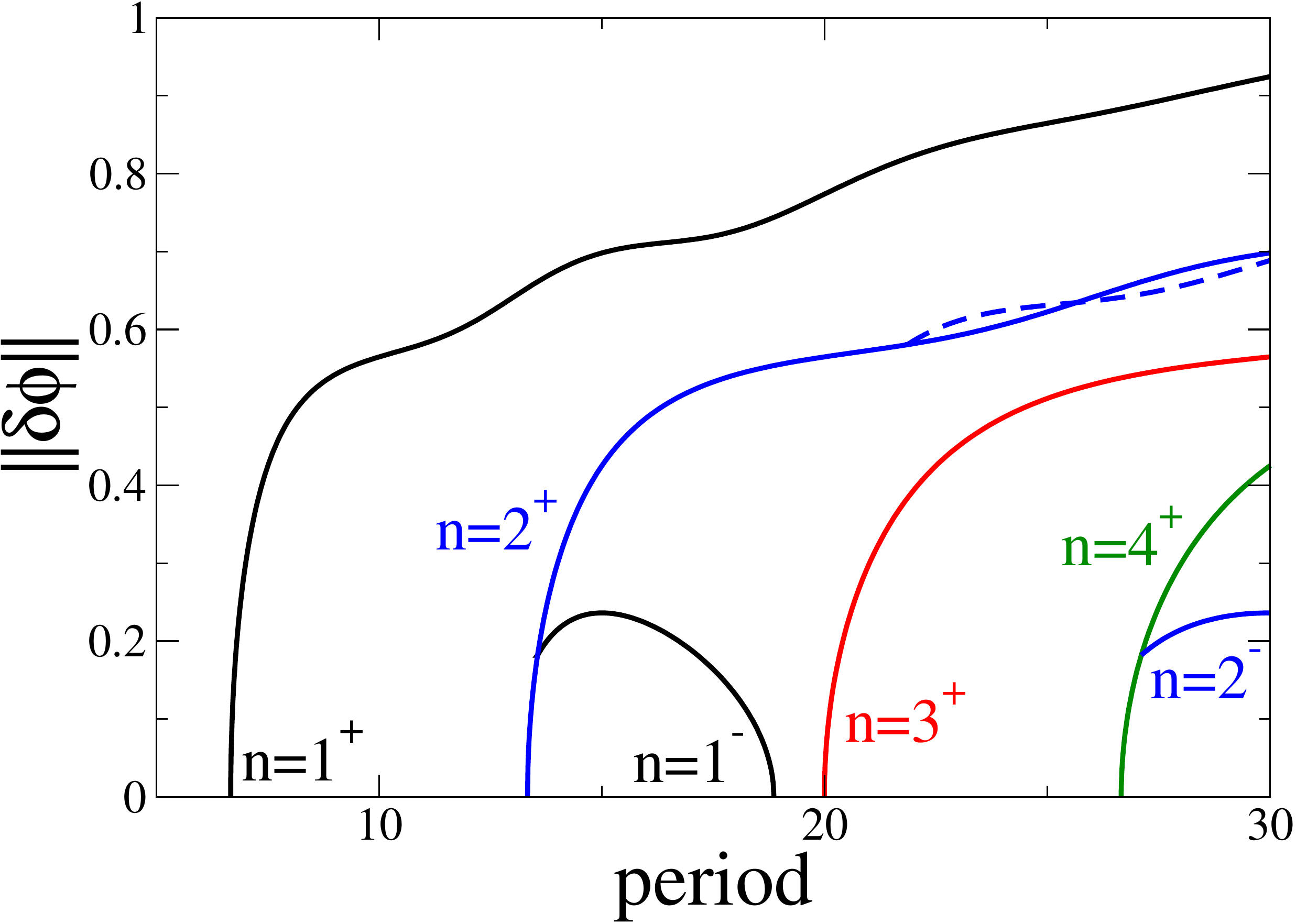}\hspace{0.05\hsize}
\includegraphics[width=0.45\hsize]{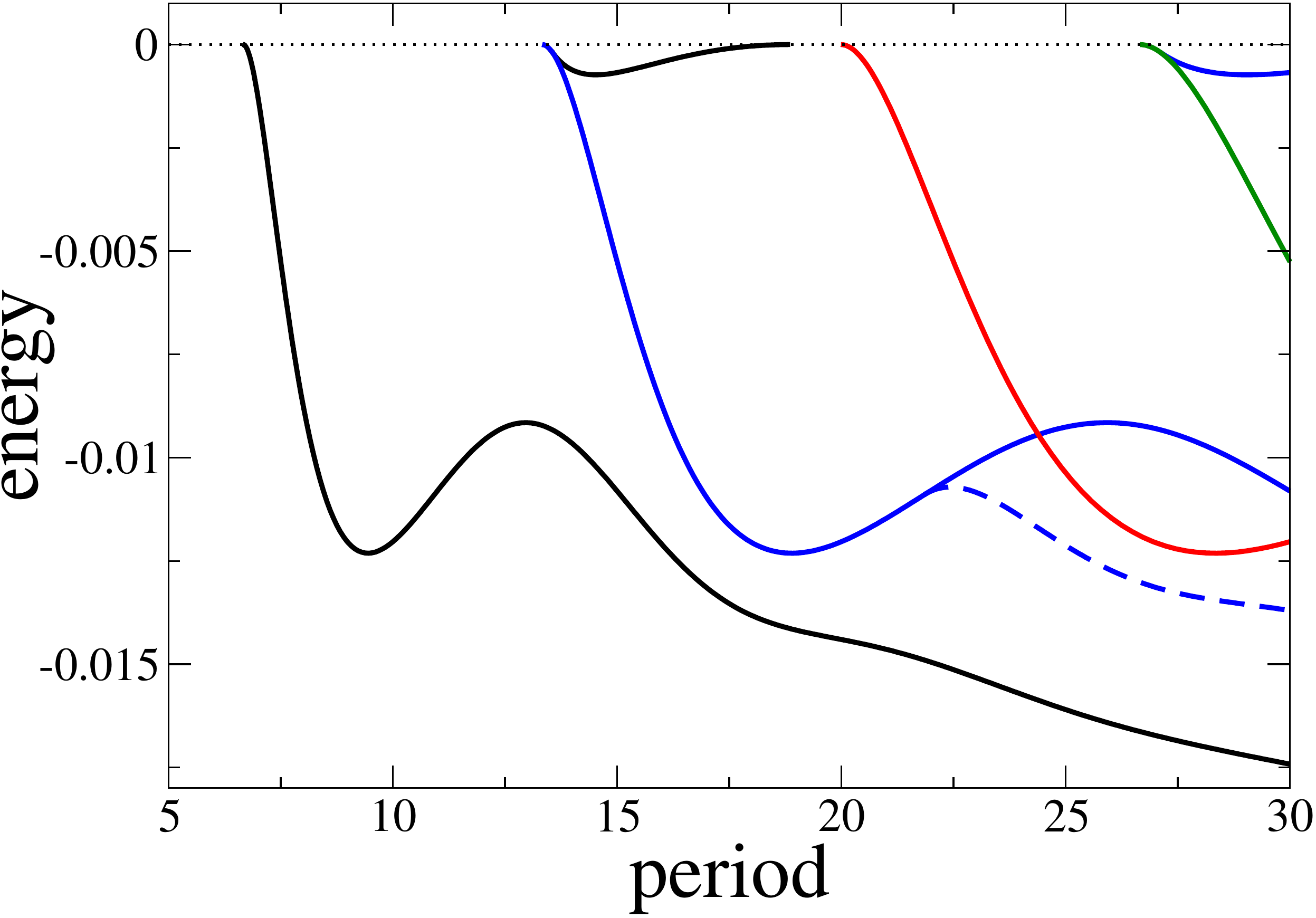}\\
\includegraphics[width=0.45\hsize]{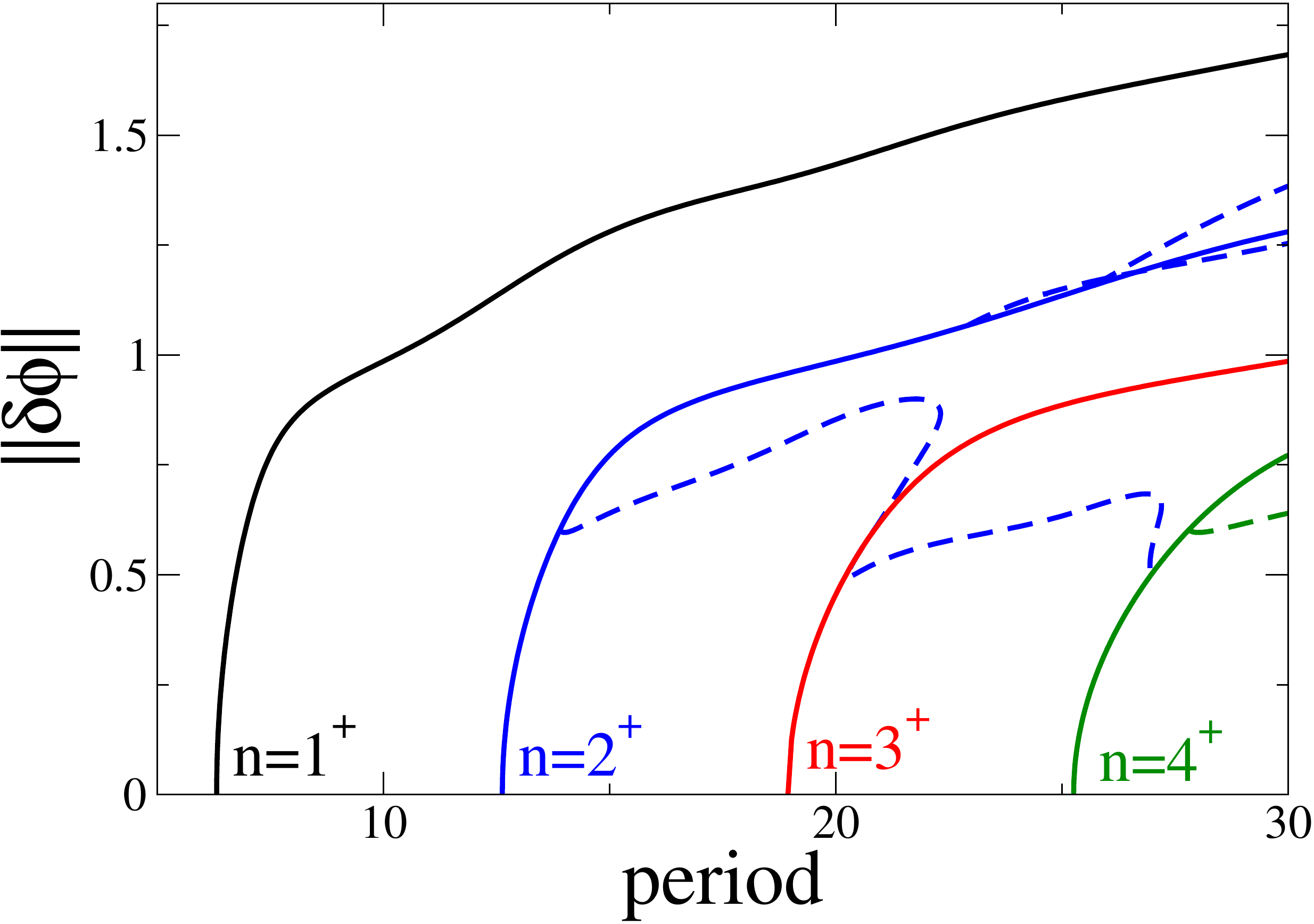}\hspace{0.05\hsize}
\includegraphics[width=0.45\hsize]{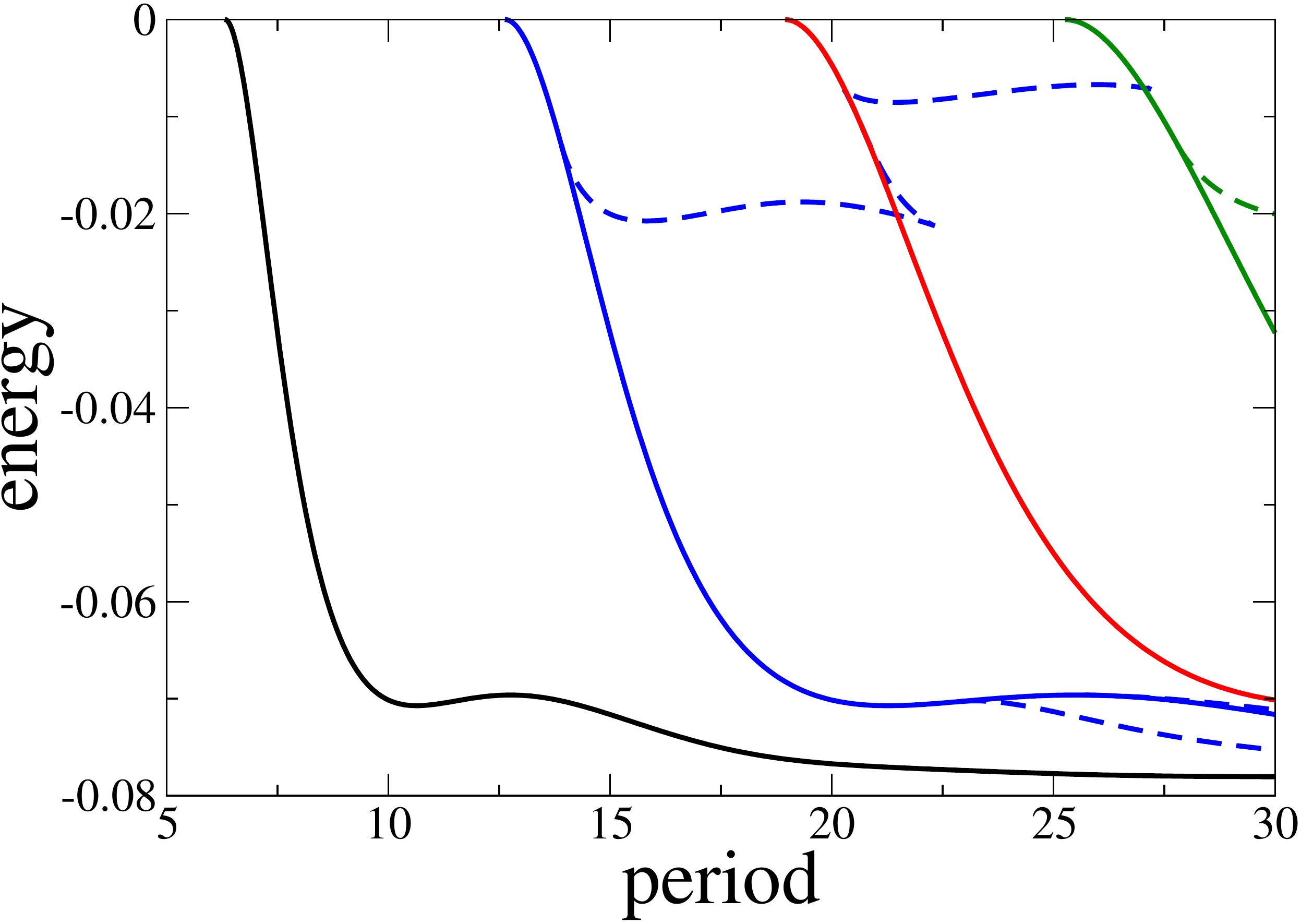}
\caption{Shown are characteristics of families of steady
  one-dimensional surface structures (quantum dots) arising in
  epitaxial growth using the simplest model proposed in
  Ref.~\cite{GLSD04}. The panels in the left column give the $L_2$-norm
  and the ones in the right column the energy $F$ (Eq.~(\ref{eq:en2}))
  per length. Parameter values correspond to (top row) Fig.~4(a),
  (middle row) Fig.~4(b), and (bottom row) Fig.~4(c) of
  \cite{GLSD04}. From top to bottom the non-dimensional wetting
  interaction increases leading to more intricate behaviour.  The
  families are obtained using continuation techniques.}
\mylab{fig:epi-fam}
\end{figure*}

\begin{figure}
\includegraphics[width=0.8\hsize]{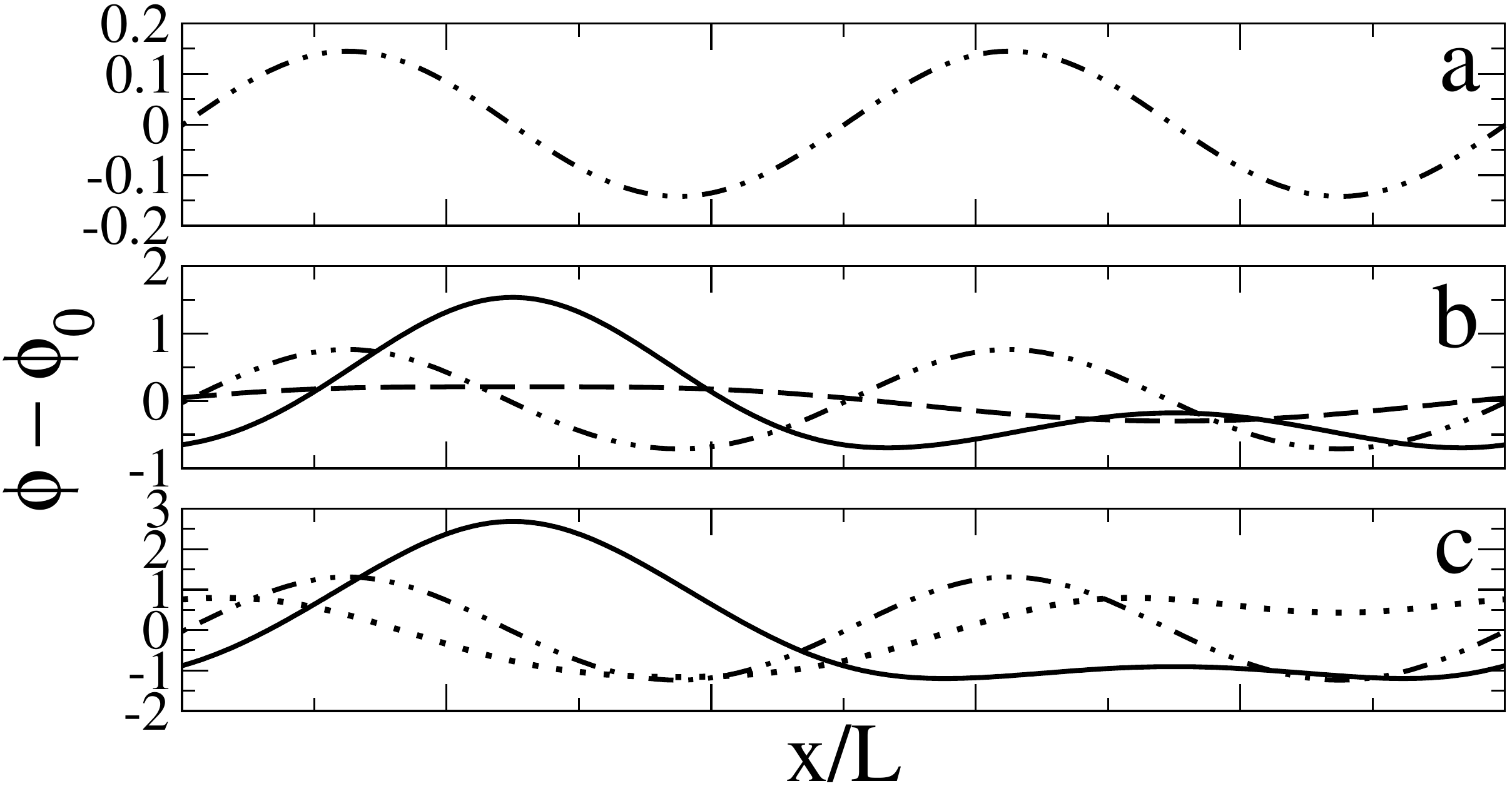}
\caption{Shown are examples of film thickness profiles ($\phi-phi_0$)
  for the various branches presented in Fig.~\ref{fig:epi-fam}.  The
  parameters correspond to (a) the top row, (b) the middle row, and
  (c) the botom row of Fig.~\ref{fig:epi-fam}.  The chosen system size
  is $L=17$ in all cases. Given are profiles from the $n=1^+$ branch
  (solid lines), the $n=2^+$ branch (dot-dashed lines), the $n=1^-$
  branch (dashed line in (b)), and the first side branch of the
  $n=2^+$ branch (dotted line in (c)) as defined in
  Fig.~\ref{fig:epi-fam}.  Mean film thicknesses is $\phi_0=1.75$.}
\mylab{fig:epi-prof}
\end{figure}

As in the other two cases steady thickness profiles are obtained by
setting $\partial_t \phi=0$ in Eq.~(\ref{film}). Here they do not
represent profiles of liquid droplets, but profiles of solid quantum-
or nanodots. Due to the higher order terms in the energy
(\ref{eq:en2}) as compared to the functional (\ref{eq:en1})
studied up to now, the dispersion curves (Fig.~\ref{fig:epi-disp}) are
qualitatively quite different from the ones for dewetting
(Fig.~\ref{fig:dew-disp}).  In consequence we expect a rather
different picture for the steady state solutions as well.  At the
parameter values allowing for two critical wavenumbers $k_{c1}$ and
$k_{c2}$, i.e.~for $\partial_{\phi\phi}f|_{\phi=\phi_0}>0$, one finds
two bifurcations from the trivial solutions at domain sizes
$2\pi/k_{c1}$ and $2\pi/k_{c2}$, respectively. Near onset one expects
the two emerging branches to interconnect, as at onset the solution
space should change only locally. Further above onset the
behaviour might, however, be dramatically different.  The two branches
bifurcating at $2\pi/k_{c1}$ and $2\pi/k_{c2}$ are the first of two
infinite series of primary solution branches that bifurcate at respective
domain sizes $L_{c1n}=2\pi n/k_{c1}$ and $L_{c2n}=2\pi n/k_{c2}$ with
$n=1,2,\dots$. This consideration is of importance here, as the
various primary branches may actually couple.

We illustrate the solution structure in Fig.~\ref{fig:epi-fam} using
parameter values as in Fig.~4 of Ref.~\cite{GLSD04}. Shown are the
norm (left column) and energy (right column) of the solutions on the
various branches for three (positive) values of the strength of the
wetting interaction.  A selection of film thickness profiles is given
in Fig.~\ref{fig:epi-prof}.
As the strength of the wetting interaction decreases the system gets
deeper into the unstable regime and the behaviour becomes more intricate. In the
upper row of Fig.~\ref{fig:epi-fam} we are slightly above onset and
the two $n=1$ branches that emerge at the two zero crossings of the
dispersion relation actually connect as expected. They are well
separated from the branches with $n>1$. Single-quantum-dot ($n=1$)
solutions only exist for a small range of periods $L_{c1}<L<L_{c2}$,
i.e., the system is not able to undergo any coarsening. This
corresponds well to the results obtained in \cite{GLSD04} using weakly
nonlinear analysis and integration in time. In the following we will
call the branch emerging at $iL_{c1}$ [$iL_{c2}$] the left [right]
$n=i$ branch.

The picture changes, however, further above onset, i.e., when
decreasing the wetting interactons (middle row of
Fig.~\ref{fig:epi-fam}). The two $n=1$ branches do not interconnect
any more. Actually, the left $n=1$ branch continues towards $L=\infty$
without any side branches. The left $n=2$ branch may change its
stability with respect to coarsening with increasing domain size.
Every time it changes stability a secondary branch bifurcates in a
period doubling bifurcation rather similar to the scenario described
for driven liquid films on inlined plates \cite{ThKn04}. The right
$n=1$ branch connects via one of the branching points to the left
$n=2$ branch. The behaviour becomes increasingly involved for higher
branch numbers.

Decreasing the wetting interaction even more brings the system further
above threshhold (bottom row of Fig.~\ref{fig:epi-fam}). As $k_{c1}$
becomes quite small the right $n=1$ branch bifurcates at very large domain
sizes. However, the first secondary branch of the left $n=2$ branch
now split up into various pieces that connect different left branches
via secondary bifurcations.

We will at present not go deeper into the analysis. In particular no
proper stability analysis for the steady states is done here and as
above for the dewetting system time evolution is not
touched. Simulations can be found, for instance, in
Refs.~\cite{Sieg97,GLSD04,Vved04,PaHu06,PaHu07,TeSp07}.  Note that the
results on stability sketched in the previous paragraph are infered
from the structure of the bifurcation diagrams alone. They need to be
scrutinized in detail as they are very important for the coarsening
behaviour.  The non-monotonous dependence of energy on system size in
Fig.~\ref{fig:epi-fam} indicates that coarsening will depend on
structure length in a non-trivial way. The long-time corsening for
epitaxial growth is discussed, e.g., in
Refs.~\cite{Sieg98,WaNo06,LeJo01}.

The results presented for the glued wetting-layer model indicate that
our knowledge about epitaxial growth would benefit from a more
detailed analysis that maps out the steady solutions and their
stability systematically for the various models proposed in the
literature. Numerical techniques are available to do this not only for
two-dimensional but as well for three-dimensional systems
\cite{BeTh08}.

\section{Conclusion}

In the present contribution we have applied a basic mathematical
analysis to three different physical systems involving solid and
liquid films at solid surfaces that may undergo a structuring process
by dewetting, evaporation/condensation or epitaxial growth,
respectively. The aim has been to highlight similarities and
differences of the three systems based on the observation that all of
them can be described using models of similar form, i.e., a time
evolution equation for a purely dissipative system without any inertia
based on a gradient dynamics that is characterized by mobility
functions and an underlying energy functional. Thereby the dynamics
might have a non-conserved and/or conserved part. Equations, like the
well known Cahn-Hilliard (purely conserved) and Allen-Cahn (purely
non-conserved) equations represent limiting cases (with respect to the
chosen dynamics) of this general evolution equation. They are normally
used with a particularly simple form of the energy functional as well.

The three physical systems discussed here all fit into the general
framework when chosing different mobilities and/or energy
functionals. Beside the mathematical similarity the systems do as well
model similar physical phenomena as the respective liquid and solid
films structure under the influence of their effective interaction
with the substrate. However, although most readers will agree
that the epitaxial growth of quantum dots is a subject of surface
science less of them might do so in the case of a dewetting liquid
film. We hope that our review contributes to the development of a more
unified view onto these processes of self-organisation at interfaces.

In particular, we have used two basic steps of mathematical analysis,
namely the study of the linear stability of homogeneous steady states,
i.e., flat films, and the mapping of non-trivial steady states, i.e.,
drop/hole (quantum dot) in dependence of system size for various
values of interaction constants and/or mean film thickness.  Our aim
has been to illustrate that the underlying solution structure might be
very complex as in the case of epitaxial growth but can be understood
better when comparing to the much simpler results for the
dewetting liquid film. We have furthermore shown that the continuation
techniques employed can shed some light on this structure in a
more convenient way than time-stepping methods. One might further
argue that understanding the solution structure of the quantum-dot
system might as well allow to predict pathways of evolution in time
for an epitaxial layer as the behaviour will ultimately be determined
by the steady solutions and their stability. This procedure was already
followed for thin liquid films \cite{TVN01,TBBB03,PBMT04}.

The usage of a general formulation like Eq.~(\ref{film}) does not only
relate seemingly not related physical systems mathematically.  It does
as well allow to discuss model extensions in a more unified
way. Taking epitaxial growth as example, the general form (\ref{film})
would propose that in the non-conserved case, i.e., when material is
deposited from the gas phase continuously the non-conserved part of
the equation should be $\,-\, M_{\mathrm{nc}}\delta F/\delta\phi$ with
$F$ given by (\ref{eq:en2}).  The constant depostion often used in the
literature would then only refer to the limit of a rather large
chemical potential. Oblique deposition of material \cite{SVD91} might
as well be modelled incorporating beside the vertical influx as well
lateral driving contributions.

The here reviewed formulation can be extended in a straightforward
way to the case of $N$ coupled fields $\vecg{\phi}=(\phi_1,
\phi_2,\dots,\phi_N)$ following a gradient dynamics that is governed
by an energy functional $F[\phi_1, \phi_2,\dots,\phi_N]$. The evolution 
equation is 
\begin{equation}
\partial_t \vecg{\phi} \,=\,
\nabla\cdot\left[\tens{M}_{\mathrm{c}}\cdot\nabla\frac{\delta F}{\delta\vecg{\phi}}\right]
\,-\, \tens{M}_{\mathrm{nc}}\cdot\frac{\delta F}{\delta\vecg{\phi}}
\mylab{filmvec}
\end{equation}
with the $\tens{M}_{\mathrm{c}}(\vecg{\phi})\ge0$ and
$\tens{M}_{\mathrm{nc}}(\vecg{\phi})\ge0$ being symmetric positive-definite
mobility matrices for the conserved and non-conserved part of the
dynamics, respectively. They are formed by $N\times N$ mobility
functions, respectively. Note that $\delta F/\delta\vecg{\phi}$
corresponds to a vector of dimension $N$. A typical example is the
evolution of a dewetting two-layer system where the $\phi_1$ and
$\phi_2$ correspond to the two film thicknesses, respectively.  The
formulation as a special case of (\ref{filmvec}) is derived from the
Navier-Stokes equations in the two liquid layers and appropriate
boundary conditions for spatially two- and three-dimensional settings
in Refs.~\cite{PBMT04,PBMT05,FiGo05,PBMT06}. The formulation in the
form of Eq.~(\ref{filmvec}) has the advantage that one can easily
check the consistencey of the mobility functions after a normally
rather involved derivation.  Furthermore one can draw on rather
general results, e.g., for the linear stability of flat film solutions
\cite{PBMT05}.  Note, that such a formulation will apply to any
multi-layer system in a relaxational setting, including films of
dielectric liquids in a capacitor \cite{BGS05,BaSh07} or multilayer
pending films under the influence of gravity. We expect it to hold as
well for the relaxation dynamics of multilayer epitaxial films.

Let us finally note that we have entirely excluded spatially
heterogeneous systems, i.e., systems described by equations of
form~(\ref{film}) where the energy depends explicitly on position.
Such systems are already treated for dewetting films
\cite{LeLi98,BDP99,KKS00,KKS01,BKTB02,TBBB03}, however, we are not
aware of any such studies for epitaxial growth.
We have as well excluded laterally driven systems described by equations like
(\ref{film}) that additionally include lateral driving forces. 
A comparison of sliding droplets on an incline
\cite{Thie02,ThKn04}, a driven (or convective) Cahn-Hilliard system
\cite{GNDZ01} and epitaxial growth under oblique deposition might be
rather interesting, for instance, in terms of coarsening behaviour
\cite{EmBr96,WORD03,ThKn04,WaNo06}.  
The combination of both effects, i.e., heterogeneous systems with
lateral driving allows to describe the depinning and stick-slip motion
of droplets \cite{ThKn06,ThKn06b}. The problems has up to now, however,
no counterpart in epitaxial growth.

\end{document}